\documentclass[12pt,letterpaper]{article}
\pdfoutput=1
\usepackage{graphicx,array}
\usepackage{color}
\usepackage{latexsym}
\usepackage{amsthm}
\usepackage{amsmath}
\usepackage{amssymb}
\usepackage{hyperref}
\usepackage{bbold}
\usepackage{subfig}

\def\simgt{\mathrel{\lower2.5pt\vbox{\lineskip=0pt\baselineskip=0pt
           \hbox{$>$}\hbox{$\sim$}}}}
\def\simlt{\mathrel{\lower2.5pt\vbox{\lineskip=0pt\baselineskip=0pt
           \hbox{$<$}\hbox{$\sim$}}}}

\newcommand{\be}{\begin{equation}}
\newcommand{\ee}{\end{equation}}
\newcommand{\bea}{\begin{eqnarray}}
\newcommand{\eea}{\end{eqnarray}}
\newcommand{\Eq}[1]{Eq.~(\ref{#1})}

\newcommand{\Fig}[1]{Fig.~(\ref{#1})}

\newcommand{\App}[1]{App.~\ref{#1}}

\newcommand{\keV}{\textrm{ keV}}
\newcommand{\MeV}{\textrm{ MeV}}
\newcommand{\GeV}{\textrm{ GeV}}
\newcommand{\TeV}{\textrm{ TeV}}
\newcommand{\gsim}{\lower.7ex\hbox{$\;\stackrel{\textstyle>}{\sim}\;$}}
\newcommand{\lsim}{\lower.7ex\hbox{$\;\stackrel{\textstyle<}{\sim}\;$}}

\hypersetup{colorlinks,citecolor= blue,linkcolor= black}

\begin{document}


\hfill

\vspace{1.5cm}

\begin{center}
{\LARGE\bf
Freeze-In Dark Matter \\ with  \vspace{.4cm}\\
Displaced Signatures at Colliders
}
\\ \vspace*{0.5cm}

\bigskip\vspace{1cm}{
{\large \mbox{Raymond T. Co, Francesco D'Eramo, Lawrence J. Hall and Duccio Pappadopulo} }
} \\[7mm]
{\it
 Berkeley Center for Theoretical Physics, Department of Physics, \\
     and Theoretical Physics Group, Lawrence Berkeley National Laboratory, \\
     University of California, Berkeley, CA 94720, USA} \end{center}
\bigskip
\centerline{\large\bf Abstract}

\vspace{0.3cm}

\begin{quote} \small

Dark matter, $X$, may be generated by new physics at the TeV scale during an early matter-dominated (MD) era that ends at temperature $T_R \ll \TeV$.  Compared to the conventional radiation-dominated (RD) results, yields from both Freeze-Out and Freeze-In processes are greatly suppressed by dilution from entropy production, making Freeze-Out less plausible while allowing successful Freeze-In with a much larger coupling strength.  Freeze-In is typically dominated by the decay of a particle $B$ of the thermal bath, $B \rightarrow X$.  For a large fraction of the relevant cosmological parameter space,  the decay rate required to produce the observed dark matter abundance leads to displaced signals at LHC and future colliders, for any $m_X$ in the range$\keV < m_X < m_B$ and for values of $m_B$ accessible to these colliders.   This result applies whether the early MD era arises after conventional inflation, when $T_R$ is the usual reheat temperature, or is a generic MD era with an alternative origin.  In the former case, if $m_X$ is sufficiently large to be measured from kinematics, the reheat temperature $T_R$ can be extracted.  Our result is independent of the particular particle physics implementation of $B \rightarrow X$, and can occur via any operator of dimension less than 8 (4) for a post-inflation (general MD) cosmology. An interesting example is provided by DFS axion theories with TeV-scale supersymmetry and axino dark matter of mass GeV to TeV, which is typically overproduced in a conventional RD cosmology.  If $B$ is the higgsino, $\tilde h$, Higgs, W and Z particles appear at the displaced decays, $\tilde h \rightarrow h \tilde a, Z \tilde a$ and $\tilde h^\pm \rightarrow W^\pm \tilde a$. The scale of axion physics, $f$, is predicted to be in the range $(3\times10^8 - 10^{12})$ GeV and, over much of this range, can be extracted from the decay length.

\end{quote}

\newpage
\tableofcontents

\section{Introduction}
\label{sec:intro}
In the early universe, when the background radiation had a temperature from eV to MeV, the energy density was dominated by radiation.  How far back did this radiation dominated (RD) era extend?  When inflation ended, provided the inflaton decays were not extremely rapid, there was an era of matter domination (MD) which ended at the reheat temperature $T_R$.  It is commonly assumed that $T_R$ is very high, many orders of magnitude above the TeV scale, but observationally the most stringent constraints are from the effective number of neutrino species from CMB measurements, $T_R > 4$ MeV \cite{Hannestad:2004px}, and from Big Bang Nucleosynthesis \cite{Kawasaki:2000en}.  This early MD era could extend for many decades in temperature above $T_R$, and include the TeV epoch.  There are several alternative origins for a long early MD era, including long-lived heavy particles that were once in thermal equilibrium and oscillating fields composed of light bosons.  When the MD era results from inflation it has an evolution that is purely Non-Adiabatic in character, giving a MD$_{NA}$ era, as shown in Fig. \ref{fig:Eras2}.   The more general MD era splits into two, beginning with Adiabatic evolution, MD$_A$, and ending with MD$_{NA}$, as illustrated in Fig. \ref{fig:Eras}.

In this paper we explore the consequences of having the dark matter abundance determined by reactions during an early MD era.  This profoundly changes the abundance compared to the conventional assumption of dark matter genesis during the RD era for two reasons.  Firstly the time-temperature relation is changed so that any given temperature occurs at earlier times and secondly, once the genesis reactions stop, the comoving abundance of dark matter is diluted by entropy production.  The second effect dominates, so that the MD era results in a {\it depletion} of dark matter relative to the usual RD result; by a factor $(T_R/T_F)^3$ for Freeze-Out during MD$_{NA}$\cite{McDonald:1989jd, Chung:1998rq, Giudice:2000ex}, where $T_F$ is the Freeze-Out temperature, and by a factor $T_R/(T_M T_F)^{1/2}$ for Freeze-Out during MD$_A$ \cite{Kamionkowski:1990ni} where $T_M$ is the temperature at which $MD_A$ begins.  This factor is likely very small, suggesting a dark matter mass much larger than the TeV scale.  In this paper we focus on TeV scale physics as the origin of dark matter, as this could be probed at colliders, and therefore we seek an alternative to Freeze-Out, as Freeze-Out during a MD era is unlikely to yield sufficient dark matter.

Particle dark matter requires an addition to the Standard Model: a cosmologically stable particle $X$ of mass $m_X$.  How is the abundance of $X$ determined?  The Freeze-Out mechanism results if $X$ has sufficient interactions with the known particles that it is in thermal equilibrium at temperatures $T$ of order $m_X$.  As $T$ drops below $m_X$,  $X$ tracks a Boltzmann distribution for a while but, as it becomes more dilute, its annihilation rate drops below the expansion rate and it freezes out of thermal equilibrium.  This mechanism has great generality, applying to a wide range of theories where the interactions of $X$ are sufficient to put it in thermal equilibrium at $T \sim m_X$.   Furthermore, it is highly predictive since it is an IR dominated process - it does not depend on any physics at energy scales above its mass.  Indeed, Freeze-Out during the RD era suggests $m_X$ is broadly of order a TeV and that $X$ has an interaction rate that can lead to signals at both collider detectors and direct and indirect detection experiments.  

Freeze-In provides another general production mechanism and results when the interactions of $X$ are insufficient to bring it into thermal equilibrium \cite{Hall:2009bx}.\footnote{We do not consider the case of dark matter arising from decay of the matter that dominates the energy density during the MD era \cite{Moroi:1999zb}.} At temperatures $T$ above $m_X$ these feeble interactions allow the production of $X$ from decays or scatterings of some bath particle $B$ of mass $m_B$ at rate $\Gamma(T)$.   This produces a yield of $X$ particles at time $t(T)$ of
\be
Y_X^{Prod}(T) \, \sim \, \Gamma(T) \; t(T)
\label{eq:YProdin}
\ee 
which is IR dominated when the interactions between $B$ and $X$ are of dimension 4 or less.  Taking $B$ to be the lightest observable sector particle carrying the stabilizing symmetry, the production rate becomes Boltzmann suppressed below $m_B$, so that the dominant contribution to $Y_X$ arises at $T \sim m_B$.  For $T>m_B$, the yield $Y(X)$ grows towards its equilibrium value, but it never reaches equilibrium, and this Freeze-In towards equilibrium stops once $T$ drops below $m_B$.  Decays generally dominate over scatterings, so that in this paper we study the reaction
\be
B \rightarrow A_{\rm SM} X
\label{eq:decay}
\ee
where $A_{\rm SM}$ is one or more Standard Model particles.

When Freeze-In occurs during the RD era, the value of the $B$ lifetime that leads to the observed dark matter abundance is of order $10^{-2} \mbox{s}$ for $m_{B,X}$ of order the weak scale.  This leads to interesting signals of long-lived stopped particles at colliders if $B$ carries electric or color charge \cite{Hall:2009bx}, but not if $B$ is neutral.  The interaction between $B$ and $X$ is described by a very small coupling of order $10^{-12}$, as shown by the blue curve of Fig. \ref{fig:FIstandard2}.  

In this paper we study Freeze-In during an early MD era, requiring $T_R$ to be less than $m_B$, and find that the $B$ lifetime, $\tau_B$, is reduced. Freeze-In via scattering during the MD$_{NA}$ era was studied in \cite{Chung:1998rq, Giudice:2000ex}, whereas Freeze-in via decays for goldstino DM was studied in Ref.\cite{Monteux:2015qqa}. The reduction in $\tau_B$ arises because the relevant time scale $t(m_B)$ that appears in (\ref{eq:YProdin}) is smaller in the MD era than in the RD era and because the produced yield $Y_X^{Prod}(T \sim m_B)$ is subsequently diluted by entropy production. Both effects require an increase of the rate $\Gamma$ appearing in (\ref{eq:YProdin}). In a model-independent analysis, we find that the required $\tau_B$ depends on two particle physics parameters $(m_B,m_X)$ and two cosmological parameters $(D,T_R)$, where $D$ appears only if the MD era has a MD$_A$ portion, in which case it represents dilution. For Freeze-In during a MD$_{NA}$ era we show our results in the two panels of Fig. \ref{fig:MasterPlotINF} for values of $m_B$ relevant for LHC and future hadron colliders, respectively. The decay length $\tau_B$ in this case is function of the reheat temperature $T_R$ only. If we also allow a MD$_A$ era then we have contours for $\tau_B$ in the $(D,T_R)$ plane, as show in Figs. \ref{fig:MasterPlots} and \ref{fig:MasterPlots2}. Over a large portion of the parameter space for Freeze-In during the MD era via (\ref{eq:decay}), these $B$ decays can be probed at colliders as they lead to exotic signals displaced from the primary vertex.  This result is remarkably general, applying to a wide range of dark matter masses, $\keV < m_X < m_B$, and a wide range of interactions that induce the decay (\ref{eq:decay})\footnote{Our results also apply to $B \rightarrow XA'$, with $A'$ containing exotic non Standard Model particles, followed by rapid decays $A' \rightarrow A_{SM}$.  The Freeze-In yield and the collider decay length are then both determined by $\Gamma(B \rightarrow XA')$.}.  For Freeze-In during the MD$_A$ era, any interaction of dimension 4 or less is sufficient, while the extreme UV-insensitivity of the MD$_{NA}$ era allows any interaction of dimension less than 8.\footnote{Our results also apply if $X$ is part of a hidden sector and FI dominates dark matter production \cite{Cheung:2010gj}.}  While the interaction is still feeble, the strength is several orders of magnitude larger than in the RD case; for example, a coupling of order $10^{-7}$ for the dimension 4 case, as in the red curve of Fig. \ref{fig:FIdilution4}.

The displaced collider signals have a size depending on the production rate of $B$ and a character depending on the identity of $A_{SM}$ and $\tau_B$.   For $B$ color and charge neutral, possibilities for $A_{SM}$ include $(h,Z, \ell^+ \ell^-, \bar{q} q, \gamma)$ giving a variety of non-prompt decays.  Using Run I data, LHC experiments have placed limits on displaced vertices \cite{Chatrchyan:2012jna,Aad:2015rba},  displaced jets \cite{CMS:2014wda,Aad:2015uaa} and displaced leptons \cite{Aad:2014yea,CMS:2014hka}.  However, some of these analyses have assumed particular forms for the signal, and may not be sensitive to, or optimized for, Freeze-In.  For example, it would be interesting to search for displaced signals involving a Higgs or $Z$ boson, as arise from $B \rightarrow hX, ZX$.  For $B$ carrying color and/or charge, $A_{SM}$ includes $(W, \ell^\pm, q, \ell \nu, \bar{q} q, g)$ giving decays from long-lived particles stopped in the detector \cite{Aad:2013gva,Khachatryan:2015jha}.

Axino dark matter in supersymmetric DFS theories of the QCD axion provides an interesting example of Freeze-In.  With Freeze-In during the conventional RD era too much dark matter is generically produced, unless theories are crafted to have very high axion scales, $f$, or very low axino masses.  However, for Freeze-In during the MD era, axinos account for the observed dark matter for $f$ in the range $(3\times10^8 - 10^{12})$ GeV and axino masses in the GeV to TeV range.  Furthermore, as shown in Figs.~(\ref{fig:MasterPlotINFAxino}) and (\ref{fig:MasterPlotsAxino}), over a large fraction of the parameter space not excluded by axion overproduction and white dwarf constraints, there are displaced signals at colliders that arises from the higgsino decays that induce the cosmological abundance, $\tilde h\to h \tilde a, \, Z \tilde a$ and $\tilde h^\pm \rightarrow W^\pm \tilde a$.

\section{An Early Matter Dominated Era}
\label{sec:Eras}
In this section we review the evolution of the universe during an early Matter Dominated era, MD, and study the parameter space of such a cosmology relevant for Freeze-In (FI) of dark matter.  We consider an FRW cosmology with two components: Radiation $( R)$ and unstable Matter $(M)$, that decays to radiation with a decay rate $\Gamma_M$.   The radiation is thermal at temperature $T$ and has energy density
\be
\rho_R \,=\, \frac{\pi^2}{30} \, g \,T^4
\label{eq:rhoR}
\ee
where $g = g(T)$ counts the effective number of relativistic species at $T$, including particles from the Standard Model and beyond.   The matter $M$ can be fermonic or bosonic and, in the era of interest to us, is not thermally coupled to the radiation.  The $M$ particles might have been in thermal equilibrium at an earlier epoch when they were relativistic, so that they have a number density comparable to that of the relativistic species, or they might arise from a classical oscillating bosonic field, as with inflatons or moduli, in which case their number densities could be much larger.  When $T$ is of order the MeV scale $M$ have decayed and the universe is in the conventional Radiation Dominated era, RD, but at earlier times we insist that the initial density of $M$ is sufficiently large that there is a long matter dominated era.  

The Boltzmann equations governing the evolution of the energy densities $\rho_{R,M}$ are 
\begin{eqnarray} 
\frac{d \rho_M}{dt} + 3 H \rho_M &=& - \Gamma_M \rho_M \nonumber \\ 
\frac{d \rho_{\rm R}}{dt} + 4 H \rho_{\rm R} &=& \Gamma_M \rho_M
\label{eq:BE}
\end{eqnarray}
where $H$ is the Hubble parameter. \footnote{The second line of Eq.~(\ref{eq:BE}) is derived from the first law of thermodynamics under the assumption that the equation of state $p_R = \rho_R / 3$ is satisfied. Strictly speaking, this is not exactly true when the temperature is near the mass of any of the particles in the thermal bath. We verified that the corrections to the equation of state are small and they only affect the final results by a few percent. Therefore, we take a constant value of $g_*$ in our numerical study. Freeze-in is dominated at temperatures of the order of the weak scale, so we take $g=106.75$, accounting for the full Standard Model degrees of freedom.}  The solutions to these Boltzmann equations depend on the initial densities $\rho_{R_i,M_i}$ and $\Gamma_M$. At times earlier than $\tau_M = 1/\Gamma_M$, the effect of decays on $\rho_M$ can be ignored, giving
\begin{eqnarray}  \rho_M &=& \rho_{M_i} \, \left(\frac{a_i}{a} \right)^3 \nonumber \\ 
 \rho_{\rm R} &=& \rho_{R_i} \, \left(\frac{a_i}{a} \right)^4 + \frac{2}{5} \left(\frac{3}{8\pi}\right)^{1/2} \Gamma_M M_{Pl} \, \rho_{M_i}^{1/2} \left(\left(\frac{a_i}{a} \right)^{3/2} - \left(\frac{a_i}{a} \right)^4 \right)
\label{eq:BEsol}
\end{eqnarray}
where $a$ is the Robertson-Walker scale factor and $M_{Pl}=G^{-1/2}=1.22\times 10^{19}$\,GeV.   There are two contributions to the radiation energy density, the first is the red-shifted initial density while the second, proportional to $\Gamma_M$,  arises from $M$ decay.  

\begin{figure}
\begin{center}
\includegraphics[height=2.6in]{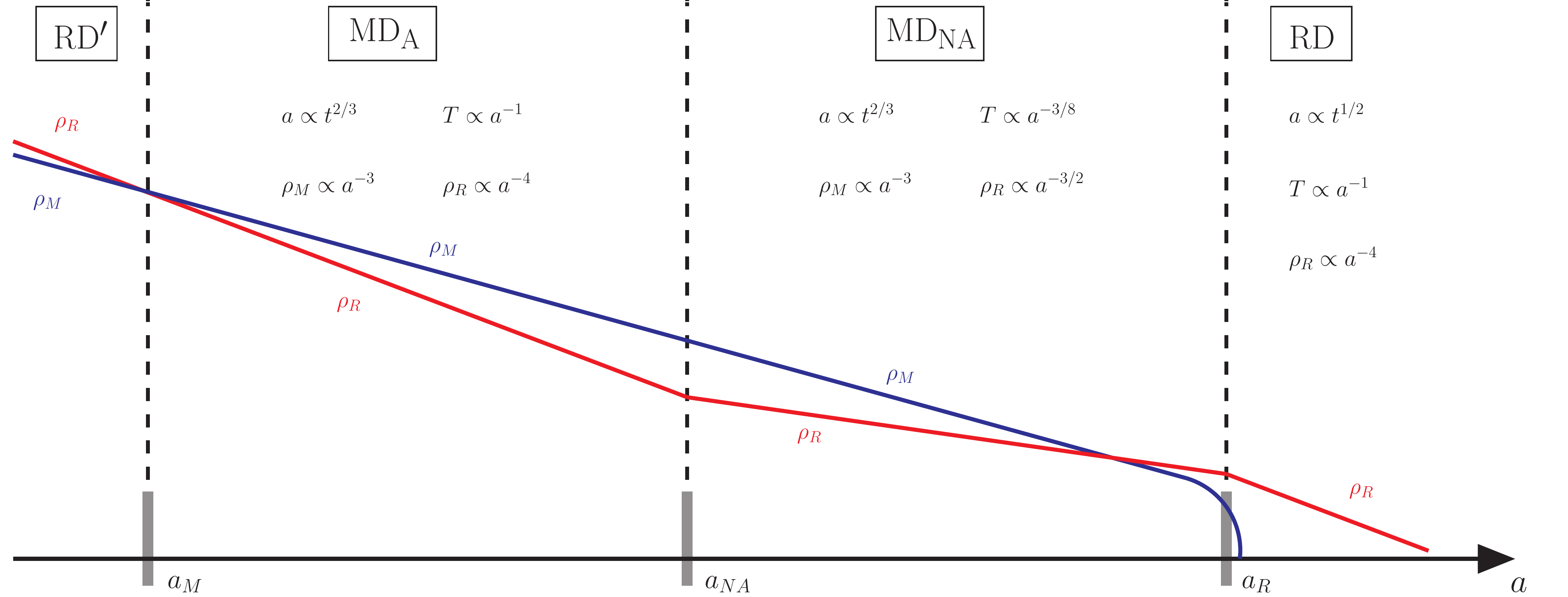}
\end{center}
\caption{Cosmological history with $a_i < a_M$. A long early MD era commences at $a_M$; it has an initial regime with adiabatic evolution and then at $a_{NA}$ it transitions to Non-Adiabatic evolution, where $\rho_R$ is dominated by decays of $M$ even though $t \ll \tau_M$.  Finally, the majority of $M$ decay at $a_R$ where the conventional RD era begins.}
\label{fig:Eras}
\end{figure}

This solution is illustrated in Figure \ref{fig:Eras} for $a_i$ sufficiently small that the universe starts in an early radiation dominated era, RD$'$. 
It transitions at $a_M$ to the early MD era, which ends at reheat, $a_R$, when $M$ decay and $\rho_M$ drops to zero.  The MD era is divided into two regimes. In the earliest one $\rho_R$ is dominated by red-shifted initial radiation, so that the second term in $\rho_R$ can be ignored and the evolution is Adiabatic (the MD$_A$ era). The second regime has $\rho_R$ dominated by radiation from $M$ decays, and hence the evolution is Non-Adiabatic (the MD$_{NA}$ era); the first term in $\rho_R$ can be ignored so $\rho_R \sim \Gamma_M M_{Pl} \, \rho_M^{1/2}$.  If $\rho_{R_i}$ is sufficiently large the universe evolves through both MD regimes; otherwise there is only a MD$_{NA}$ regime. The transition between these two types of cosmology occurs at
\be
\rho_{R_i} \sim (\Gamma_M M_{Pl}) \, \rho_{M_i}^{1/2}.
\label{eq:NAtransition}
\ee

\subsection{General MD Histories Relevant for Freeze-In}
\label{subsec:genhist}
In this paper we study Freeze-In of dark matter during this MD era.  Taking $a_i = a_M$ matches on smoothly to the RD$'$ era, but it could be that the Boltzmann equations of (\ref{eq:BEsol}) break down at some $a$ larger than $a_M$, so that the MD era emerges from a non-RD era, such as inflation or topological defect domination.  To be completely general we allow $a_i$ to be anywhere between $a_M$ and $a_R$.  (Taking $a_i \ll a_M$ gives an identical MD era to the case of $a_i = a_M$, and $a_i>a_R$ does not have an early MD era.) Equivalently, we study all initial conditions with
\be
\rho_{M_i} \geq \rho_{R_i}, (\Gamma_M M_{Pl})^2.
\label{eq:IC}
\ee
In fact we insist that $\rho_{M_i} \gg (\Gamma_M M_{Pl})^2$ so that the MD era is not a transient phenomenon.  

\begin{figure}
\begin{center}
\includegraphics[scale=0.49]{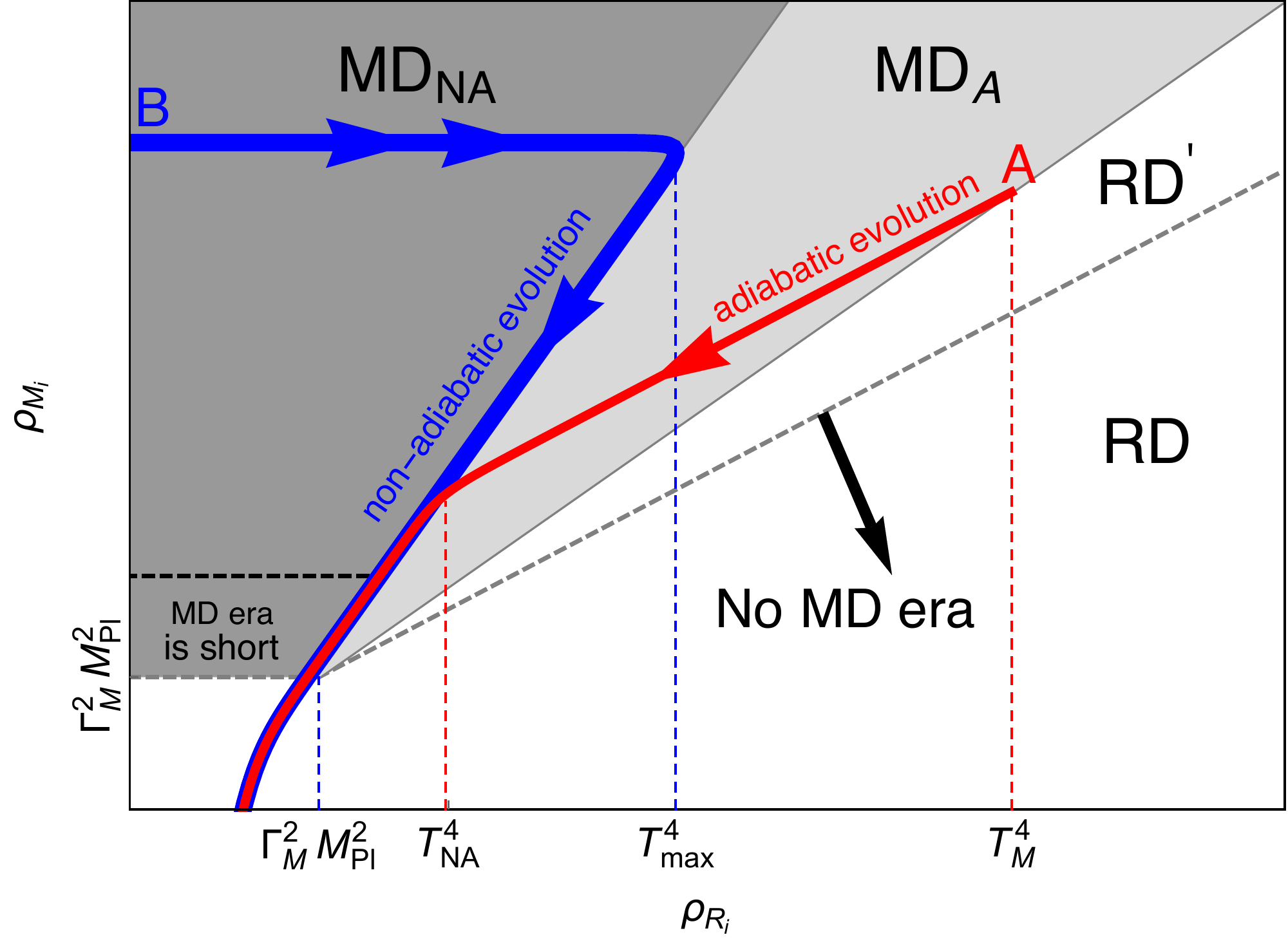}
\end{center}
\caption{The space of cosmological histories in the $(\rho_{R_i},\rho_{M_i})$ plane.  The unshaded region below the dashed line has no early MD era.  The three regions above the dashed line are distinguished by the initial era.  The blue and red trajectories give the evolution of energy densities for histories labelled A and B (with axes interpreted as $\rho_R(a)$ and $\rho_M(a)$, $a>a_i$, and several key temperatures indicated). Note that $\Gamma_M^2 M_{Pl}^2 \sim T_R^4$.}
\label{fig:Histories}
\end{figure}
The set of cosmological histories of interest to us is shown shaded in Figure \ref{fig:Histories}, and splits into two according to whether the evolution is initially MD$_A$ (light gray) or MD$_{NA}$ (dark gray).  The region above the dashed line which is initially RD$'$ will evolve to give the same MD era as histories on the boundary between initially RD$'$ and initially MD$_A$.  Two particular histories are illustrated by the red and blue trajectories having initial energy densities A and B.  Point A has $\rho_{R_i} = \rho_{M_i} \gg (\Gamma_M M_{Pl})^2$ and corresponds to $a_i = a_M$ in Figure \ref{fig:Eras}, leading to both Adiabatic and Non-Adiabatic evolution. Point B has $\rho_{R_i}=0$.  In this case, those $M$ that decay much earlier than $\tau_M$ rapidly increase the radiation density to $\rho_R \sim \Gamma_M M_{Pl} \, \rho_M^{1/2}$, leading to a MD era that is purely Non-Adiabatic.  The horizontal part of the blue trajectory is covered very quickly so that the MD$_{NA}$ evolution occurs on the non-horizontal part of the blue trajectory. For fixed $\Gamma_M$, the possible histories with purely Non-Adiabatic evolution can be described by $\rho_{M_i}$, which labels points along the line of non-adiabatic evolution.  Histories having both evolutions are described by $(\rho_{M_i}, \rho_{R_i})$, corresponding to points in the region with light shading.

\begin{figure}
\begin{center}
\includegraphics[height=2.6in]{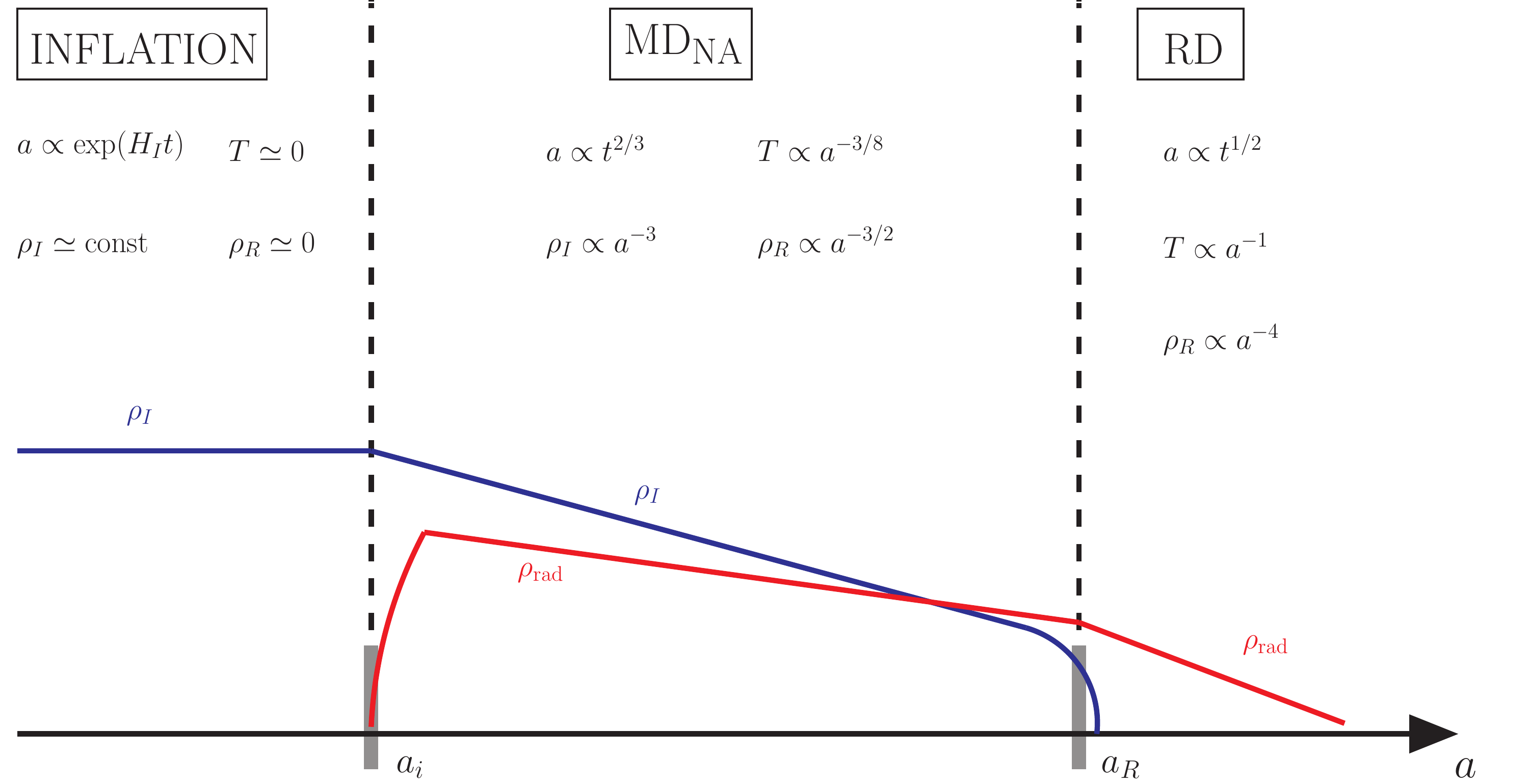}
\end{center}
\caption{A long MD era after inflation.}
\label{fig:Eras2}
\end{figure}

Histories with $\rho_{R_i}=0$, such as the blue trajectory in Fig. \ref{fig:Histories}, could emerge from an inflationary era with $M$ interpreted as the inflaton, which starts to oscillate at $a_i$, as illustrated in Fig. \ref{fig:Eras2}.

\subsection{Evolution of time, temperature and energy density}
\label{subsec:tTrho}

For the purposes of studying Freeze-In during the MD era, we lose no generality in considering histories with $\rho_{M_i} = \rho_{R_i}$ (corresponding to $a_i = a_M$, such as point A in Figure 2) since, in the late part of their evolution, they include all other histories that have larger $a_i$. In particular, solutions with equal initial conditions always contain a 
MD$_{NA}$ phase at late times; therefore they also cover histories with only a MD$_{NA}$ era (up to the very quick beginning of it, shown on the horizontal blue line in Fig.~(\ref{fig:Histories})). Hence the three dimensional parameter space ($\rho_{R_i}, \rho_{M_i}, \Gamma_M$) may be collapsed to two ($\rho_{R_i} = \rho_{M_i}, \Gamma_M$). The evolution of time and temperature in these histories, relevant for physical processes, is discussed below and the two parameters may be taken as $(T_M, T_R)$, the temperatures at the beginning and end of the MD era.  They are related to ($\rho_{R_i}, \Gamma_M$) by $\rho_{R_i} \sim T_M^4$ and $T_R \sim \sqrt{\Gamma_M M_{Pl}}$.

During the entire MD era the age of the universe is $t = t_M(a/a_M)^{3/2}$. During the MD$_A$ regime $g^{1/3}aT$ is constant. For a simple pedagogical discussion, we will assume a constant $g$ so that the energy density is
\be
\rho_A(T) \; =\; \frac{\pi^2}{30} g \left( T^4 +  T_M T^3 \right)
\label{eq:rhoA}
\ee
and in this regime the time-temperature relation has the parametric form
\be
t_A(T) \; \sim \; \frac{1}{g^{1/2}} \;\; \frac{M_{Pl}}{T_M^{1/2} \; T^{3/2}}.
\label{eq:tT-A}
\ee

The universe transitions from adiabatic to non-adiabatic evolution when the two terms in $\rho_R$ of ({\ref{eq:BEsol}}) are comparable, at time\footnote{The $\simeq$ symbol indicates approximations resulting from complicated behaviors during transition regions between eras, while the $\sim$ symbol indicates that, in addition, numerical constants have been dropped.}
\be
t_{NA} \; \sim \; t_M^{2/5} \, t_R^{3/5}
\label{eq:tNA}
\ee
with $t_M = t_A(T_M)$ and $t_R = \tau_M = \Gamma_M^{-1}$, and at temperature
\be
T_{NA} \; \sim \;  \; T_M \; \left( \frac{t_M}{t_{NA}}\right)^{2/3}.
\ee
In Fig. \ref{fig:Histories}, points along the boundary separating the initially MD$_A$ and MD$_{NA}$ regions can be labeled by $T_{NA}$.  For histories resulting from inflation there is no MD$_A$ era, rather the history begins at $T_{NA}$, which is identified with
\be
T_{max} \sim \sqrt{E_I T_R}
\label{eq:Tmax}
\ee
where $E_I$ is the energy scale of inflation.  For this case it is convenient to take the independent variable as $T_{max}$ instead of $T_M$.

During the MD$_{NA}$ regime $a^{3/8}\; T$ is constant so that the energy density is
\be
\rho_{NA}(T) \; =\; \frac{\pi^2}{30} g \left( T^4 +  \frac{T^8}{T_R^4} \right)
\label{eq:rhoNA}
\ee
where $\rho_M$ has been normalized so that it is equal to $\rho_R$ at $T_R$. This gives a time temperature relation
\be
t_{NA}(T) \; \sim \; \frac{M_{Pl}T_R^2}{g^{1/2} \, T^4}.
\label{eq:tT-NA}
\ee
Matching (\ref{eq:rhoA}) and (\ref{eq:rhoNA}) gives
\be
T_{NA}^5 \; \simeq  \; T_M T_R^4.
\label{eq:TNA}
\ee

The final RD era is reached when $T=T_R$ and $t=t_R$.  Including a numerical factor, (\ref{eq:tT-NA}) at this transition gives
\be
T_R \; \simeq \; \left( \frac{90}{8 \pi^3 g} \right)^{1/4}  \; \sqrt{\Gamma_M M_{Pl}}.
\label{eq:TR}
\ee

Above $T_{eq} \sim 1 $eV the universe was radiation dominated and it is frequently assumed that this RD era  began at extremely high temperatures, as high as the energy scale of conventional inflation, $E_I$,  which could be as large as $10^{16}$ GeV.  However, in the cosmology discussed here, the universe had a long MD era from $T_R$ to $T_M$.  Observational constraints are very mild: $T_R \gsim 3$ MeV from CMB and BBN and $T_M \lsim E_I$. Even in the case that this early MD era arose from the conventional inflaton, and hence was purely MD$_{NA}$, it could have lasted for many decades in temperature from $T_R$ as low as 3 MeV to $T_{max} \sim \sqrt{E_I T_R}$.   If the DM abundance was determined during the MD era we expect significant changes in predictions compared to the RD case for two reasons.  The DM yield depends on the Hubble parameter at production ($Y \propto H$ for FO and $Y \propto 1/H$ for FI) and at any temperature in the MD era H is larger than at the same temperature in the RD era by an amount that can be as large as $(T_M/T_R)^{2/5}$ when $T=T_{NA}$.  However this is overwhelmed by dilution effects subsequent to DM production, which reduces the yield by as much as $T_R/T_M$.  For the production of TeV-scale dark matter in a long MD era this is typically catastrophic for FO since the abundance becomes too low, but for FI it has the interesting feature of enhancing the required couplings of the weakly interacting state to yield displaced signals at colliders.


\section{Freeze-In During an Early MD Era}
\label{sec:FI}

We consider Freeze-In (FI) production of dark matter during the early MD era via the decay $B \rightarrow A_{\rm SM} X$. $B$ is the lightest observable sector particle that carries the stabilizing symmetry,  $X$ is the dark matter and $A_{\rm SM}$ is one or more Standard Model particles.   In addition to FI, there is a Freeze-Out (FO) population of $B$ that eventually decays to $X$. In this Section we focus on FI, deferring FO to Appendix \ref{sec:FO} where we show that in the early MD era it is typically negligible.

The number density evolution of $X$ is described by the Boltzmann equation~\cite{Hall:2009bx}
\be
\frac{d n_X}{dt} + 3 H n_X =  \Gamma_B  \, n_B^{\rm eq} \,\frac{K_1[m_B/T]}{K_2[m_B/T]} \ ,
\label{eq:BoltzFI}
\ee
with $\Gamma_B$ the width of $B$ and $K_{1,2}[x]$ the first and second modified Bessel functions of the 2nd kind. The equilibrium number density obtained using Maxwell-Boltzmann statistics reads
\be
n_B^{\rm eq} = \frac{g_B}{2 \pi^2} \, m_B^2 T \, K_2[m_B/T] \ .
\ee
At high temperatures ($T \gg m_B$) we recover the $T^3$ abundance for a relativistic species, whereas at low temperatures ($T \ll m_B$) the number density has the Maxwell-Boltzmann exponential suppression. For this reason, the FI production of $X$ is dominated at temperatures $T_{\rm FI} \simeq m_B$.

If FI occurs during the RD era, the Boltzmann equation~(\ref{eq:BoltzFI}) can be easily solved~\cite{Hall:2009bx} giving a final yield for $X$
\be
Y_X \, = \, \frac{n_X}{s} \, = \, 4.4\times10^{-12} \, \left( \frac{g_B}{2} \right)  \left(\frac{106.75}{g_*}\right)^{3/2}
 \left(\frac{300 \, \GeV}{m_B}\right) \left(\frac{\Gamma_B / m_B}{1.8 \times 10^{-25}}\right) \ 
\label{eq:RDFIresult}
\ee
where $s$ is the entropy density.  Observations fix the DM density but the yield is fixed once the mass of $X$ is known using
\be
\xi_{DM}=\frac{\rho_{DM}}{s}=m_{DM} Y_{DM}=0.44 {\textrm{eV}}
\ee
which is close to the usual temperature of matter radiation equality, $T_{eq} \simeq 1{\textrm{eV}}$. We can thus rewrite eq.~(\ref{eq:RDFIresult}) as
\be
\xi_{X}=\xi_{DM}\left( \frac{g_B}{2} \right)  \left(\frac{106.75}{g_*}\right)^{3/2}
 \left(\frac{m_X}{100 \, \GeV}\right) \left(\frac{300 \, \GeV}{m_B}\right) \left(\frac{\Gamma_B / m_B}{1.8 \times 10^{-25}}\right).
 \label{eq:xiX}
\ee
The coupling $\lambda$, defined by 
\be
\Gamma_B = \frac{\lambda^2}{8 \pi}  m_B,
\label{eq:FIparameters}
\ee
must be very small to avoid overclosure. For the reference masses and number of spin states shown in (\ref{eq:xiX}) the observed DM density results for $\lambda \simeq 2 \times 10^{-12}$. 

\begin{figure}
\begin{center}
\includegraphics[scale=0.55]{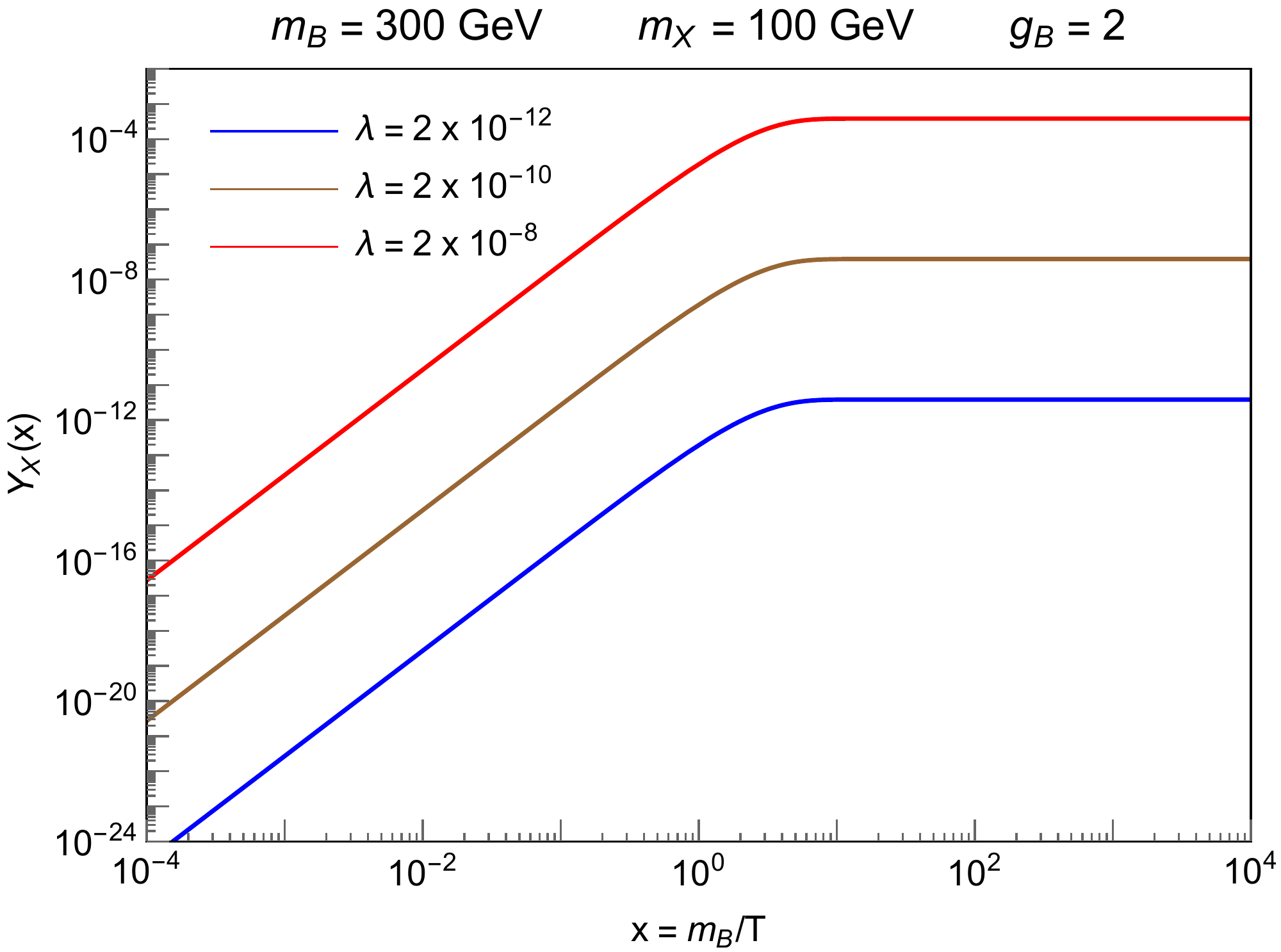}
\end{center}
\caption{Yields $Y_X(x)$ for FI during the RD era for three values of the decay width, as defined in \Eq{eq:FIparameters}, with fixed reference values $m_B = 300 \, {\rm GeV}$, $m_X = 100 \, {\rm GeV}$, and $g_B = 2$.}
\label{fig:FIstandard2}
\end{figure}

In \Fig{fig:FIstandard2} we show the FI solutions for the DM yield $Y_X(x)$, with time variable $x = m_B / T$. We show three different lines corresponding to three different values of $\lambda$ (namely three different decay widths as in \Eq{eq:FIparameters}). In all cases we see that the process is maximally active at temperatures of order $m_B$.   For the reference parameters of (\ref{eq:RDFIresult}), the blue line reproduces the observed DM density, whereas the brown and red lines overproduce the amount of dark matter. In particular, the red line reaches an asymptotic value of $Y_X$ which is $10 \%$ the equilibrium value $Y_X^{\rm eq}$, and above this point the approximation that DM arises from FI through the decay in \Eq{eq:decay}, and without the inverse reaction, breaks down.

The FI abundance depends on the cosmological evolution at and after FI.  The solution of the Boltzmann equation (\ref{eq:BoltzFI}) is given in Appendix \ref{sec:FIsol} for an arbitrary cosmological evolution in Eqs. (\ref{eq:FIandDilIntegral}) and (\ref{eq:H}).  In this section we apply this to the cosmological backgrounds discussed in the previous section, with energy density $\rho_R + \rho_M$ evolving according to the Boltzmann equation (\ref{eq:BE}).  These cosmological backgrounds depend in general on three parameters, $\rho_{R_i}, \rho_{M_i}$ and $\Gamma_M$ (or equivalently $T_R$).  In the next subsection we study the case of FI during the MD$_{NA}$ era, which is particularly simple because there is no dependence on $\rho_{R_i}$ or $\rho_{M_i}$ and the results can be displayed as a function of $T_R$.

\subsection{Freeze-In during Non-Adiabatic evolution}
\label{subsec:FINA}

Freeze-In during MD$_{NA}$ occurs between $a_{NA}$ and $a_R$, as shown in Fig.~\ref{fig:Eras}, and the final DM density depends on only one parameter of the background, the reheat temperature $T_R$. This includes the important example of FI during reheating after inflation, as sketched in Fig.~\ref{fig:Eras2}.  In this case inflation could be the conventional one with more than 60 e-foldings with the matter M identified as the usual inflaton, or it could be a later era of inflation with fewer e-foldings.  
 \begin{figure}
\begin{center}
\includegraphics[scale=0.55]{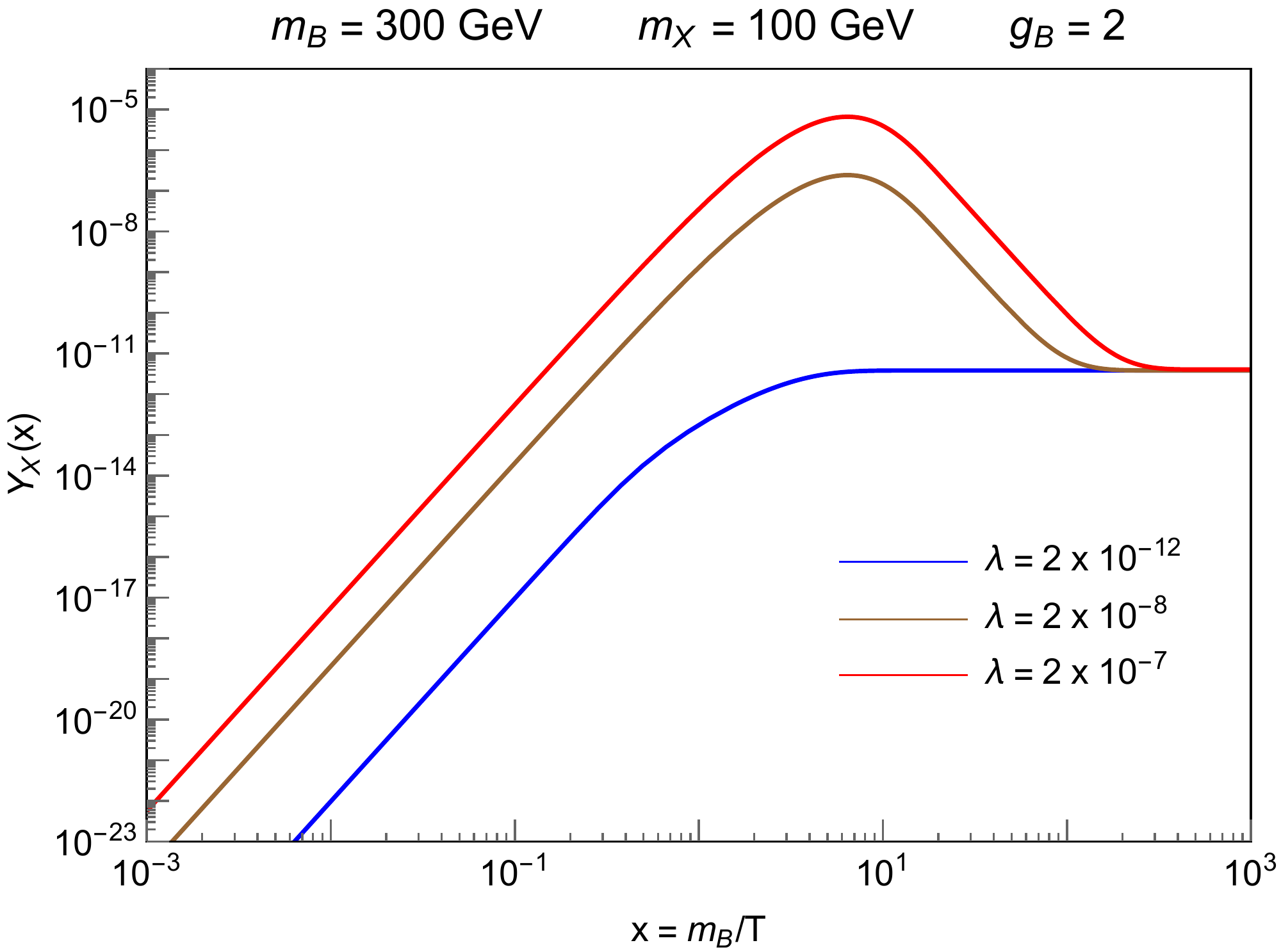}
\end{center}
\caption{Yield $Y_X(x)$ for Freeze-In during the MD$_{NA}$ era, for example during reheating after inflation.  For each of three $\lambda$, $T_R$ is chosen to give the observed DM abundance, as shown in  \Eq{eq:FIparameters4}. We take fixed reference values $m_B = 300 \, {\rm GeV}$, $m_X = 100 \, {\rm GeV}$, and $g_B = 2$.}
\label{fig:FIdilution4}
\end{figure}

The FI yields $Y_X(x)$ are shown in Fig.~\ref{fig:FIdilution4} for three different values of the decay width
\be
\lambda = \left\{ \begin{array}{lccl} 
2 \times 10^{-12}  & & & T_R = 1 \, \TeV   \\
2 \times 10^{-8}  & & &  T_R = 4.6 \, \GeV   \\
2 \times 10^{-7}  & & &  T_R = 2.4 \, \GeV   
\end{array} \right.  \ .
\label{eq:FIparameters4}
\ee
In each case we choose $T_R$ in such a way that we reproduce the observed DM abundance. The values of $\lambda$ span a wider range than the case of Fig.~\ref{fig:FIstandard2} since Freeze-In during a MD era allows for larger couplings. The blue line has $T_R > m_B$, so that FI occurs during the RD era.  As $\lambda$ is increased, the FI process becomes more powerful, as illustrated by the brown and red lines, and to obtain the observed DM abundance much lower values of $T_R$ must been taken, so that the $X$ abundance is diluted by entropy production after FI.
However, it is not possible to arbitrarily increase $\lambda$ and still be in the Freeze-In regime. For sufficiently large $\lambda$, the peak of the $Y_X$ functions in Fig.~\ref{fig:FIdilution4} will reach the equilibrium value and FI is no longer the applicable production mechanism.  For reference values ($m_B = 300 \, {\rm GeV}, m_X = 100 \, {\rm GeV}, g_B = 2$) this gives a lower limit $T_R \geq 0.6 \, {\rm GeV}$. At the same time we have to keep $T_R \leq m_B$; otherwise FI occurs during the RD era and there would be no dilution.

\begin{figure}[ht]
\begin{center}
\includegraphics[scale=0.38]{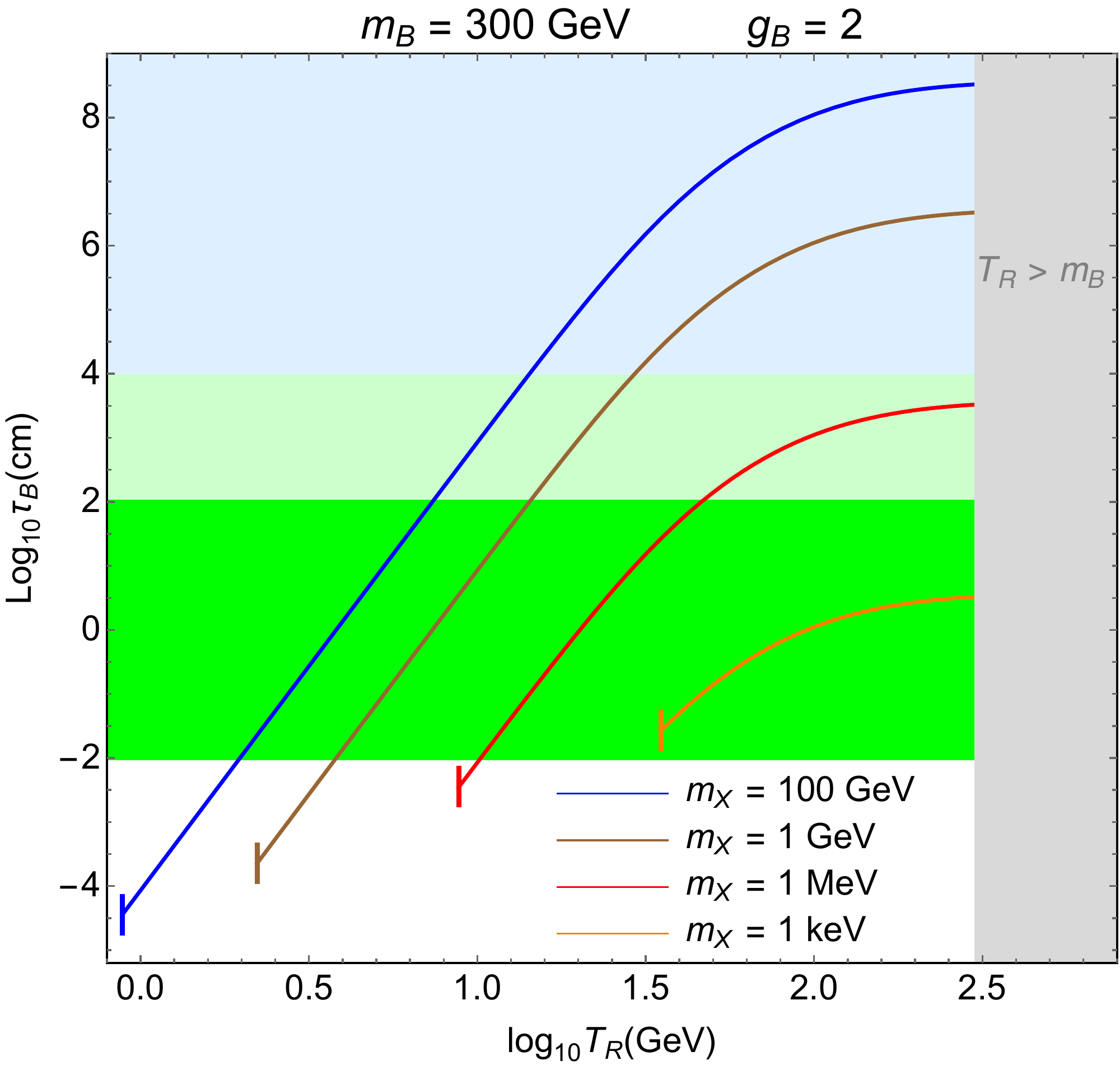} $\qquad$
\includegraphics[scale=0.38]{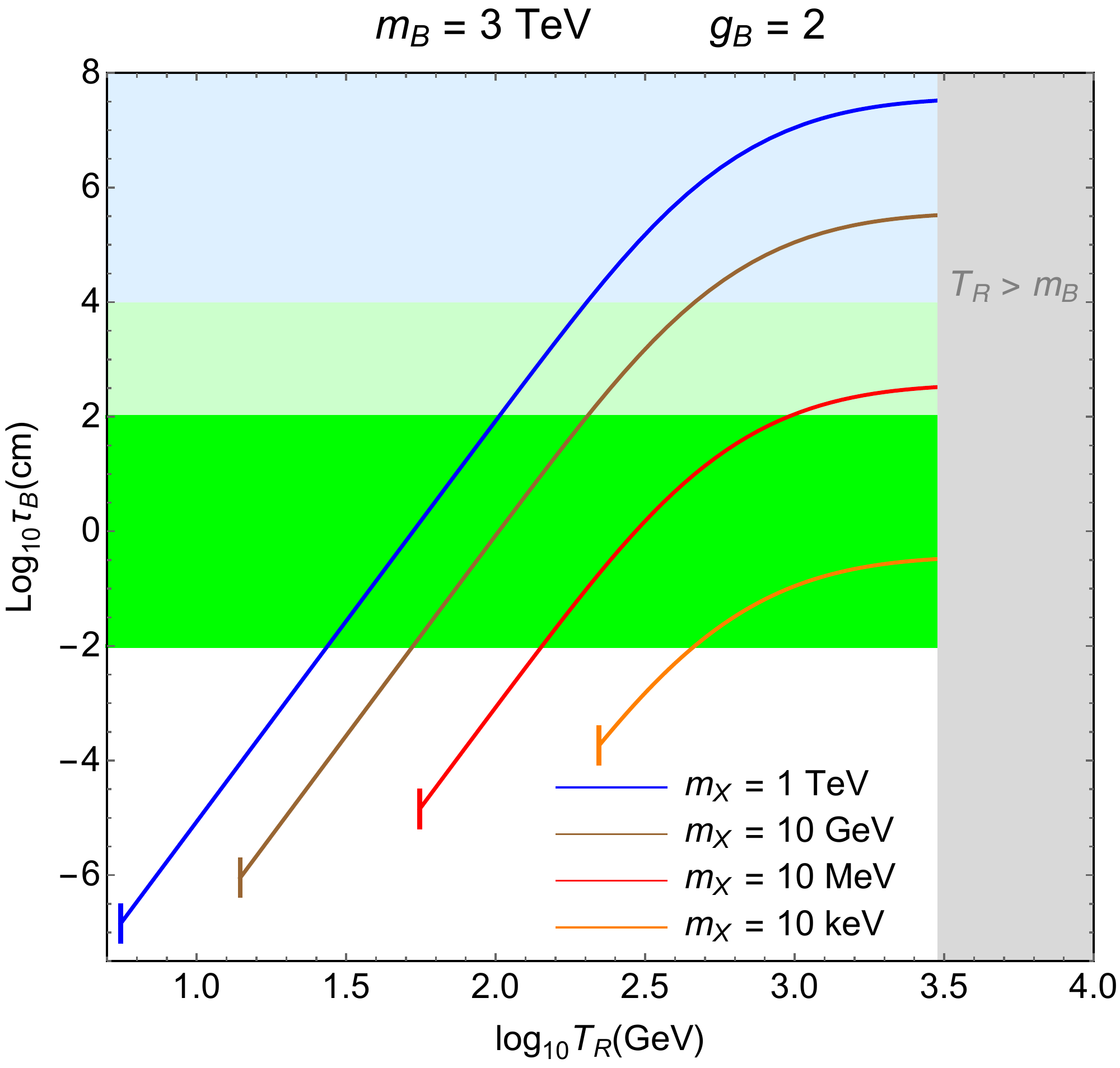} 
\end{center}
\caption{$B$ decay length as a function of $T_R$ for FI during the MD$_{NA}$ era, as occurs after inflation, for two different choices of $m_{B}$. In each panel we consider four different values of $m_X$. The various curves stop on the left because Freeze-In no longer occurs as $X$ thermalizes and on the right because Freeze-In occurs during the regular RD era.  Displaced collider signals occur in the shaded horizontal bands as described in Table~\ref{table:DisplacedColliderSignals}.}
\label{fig:MasterPlotINF}
\end{figure}

The $B$ decay length necessary to reproduce the observed DM abundance is shown in Fig.~\ref{fig:MasterPlotINF} as a function of $T_R$. The left (right) panel is for a value of $m_B = 300 \GeV$ ($3 \TeV$) relevant for the LHC (future colliders) and four curves are shown in each, giving results for a wide range of $m_X$.  Signal regions for various displaced events at colliders are given by shaded horizontal bands, as described in Table~\ref{table:DisplacedColliderSignals}.  The stopped particle decay signal is absent if $B$ is color and electric charge neutral. 

\begin{table}[hb]
\begin{center}
\begin{tabular}{| l c c c r |}
 \hline
  Shaded region & Decay length & Signature from LOSP & Neutral & Charged \\  \hline \hline
\rule{0pt}{2.3ex}Dark green  & $10^{-2} \mbox{cm} < \tau_B < 10^2 \mbox{cm} $ & \mbox{Displaced vertices}  & $\checkmark$ & $\checkmark$ \\ \hline
  Light green & $10^{2} \mbox{cm} < \tau_B < 10^4 \mbox{cm} $ & \mbox{Displaced jets/leptons}  & $\checkmark$  & $\checkmark$ \\ \hline
  Light blue & $10^4 \mbox{cm} < \tau_B $ & \mbox{Stopped particle decays}  & \sffamily X  & $\checkmark$\\ \hline
\end{tabular}
\caption{Displaced Collider Signals}
\label{table:DisplacedColliderSignals}
\end{center}
\end{table}

\subsection{Freeze-In during Adiabatic or Non-Adiabatic evolution}
\label{subsec:FIA}

We extend the previous discussion to the more general framework of Fig.~\ref{fig:Eras}, which also allows for a period of adiabatic evolution during the MD era.  The full parameter space of the cosmological background is ($\rho_{R_i}, \rho_{M_i}, \Gamma_M$) but, as discussed in Section \ref{sec:Eras}, for FI there is no loss of generality in taking $\rho_{R_i}= \rho_{M_i}$.  The parameter space may then be taken as $(T_M, T_R)$, where $T_M$ is the temperature at the beginning of the MD era. We trade $T_M$ for another parameter, the dilution $D \equiv S_f / S_i \simeq T_M / T_R$, where $S_i$ and $S_f$ are the entropy per comoving volume before and after the matter decay. 

\begin{figure}
\begin{center}
\includegraphics[scale=0.55]{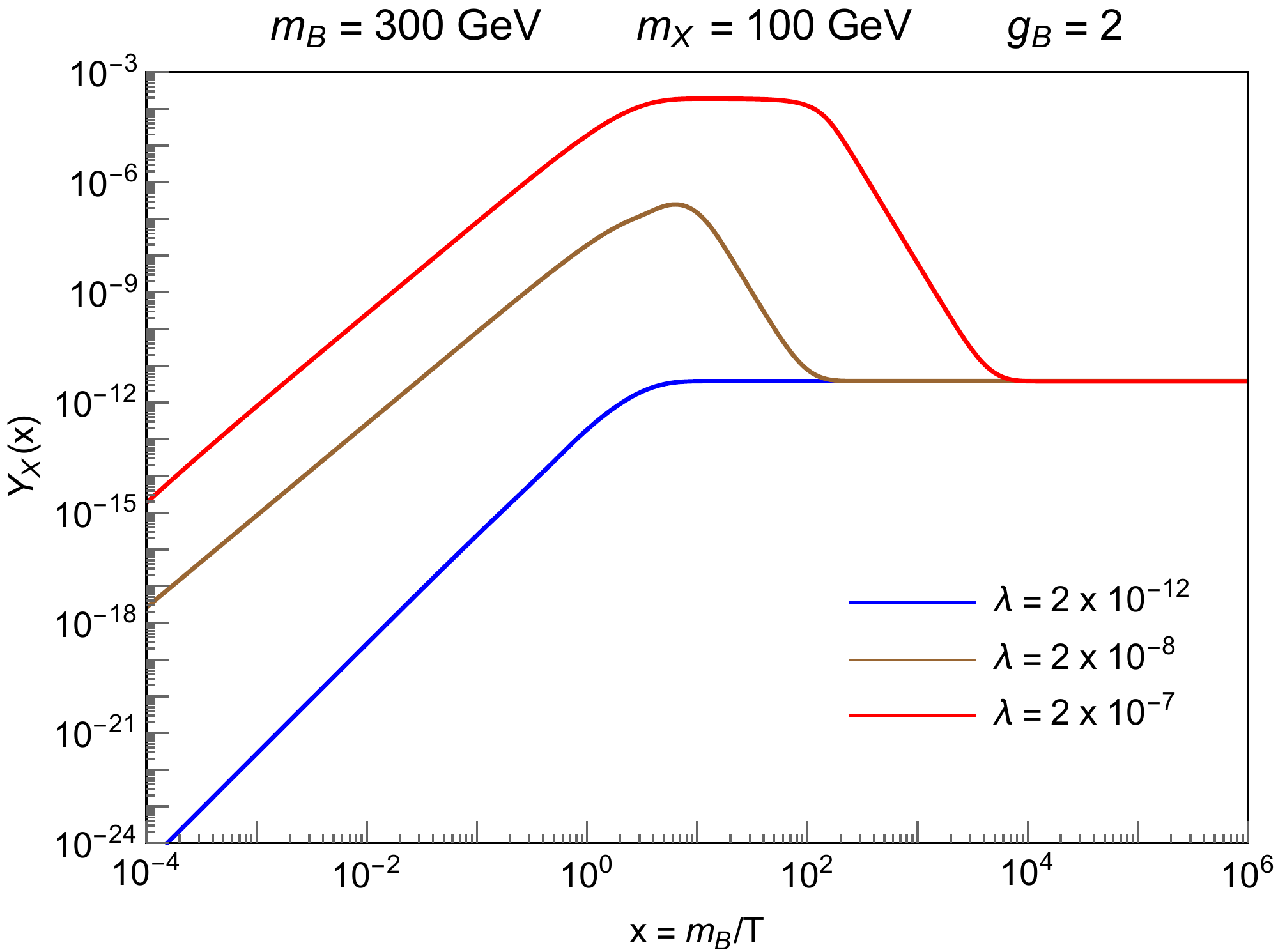}
\end{center}
\caption{DM yield $Y_X = n_X / s$ as a function of $x = m_B / T$ for Freeze-In during an early matter dominated era. For each of the three $\lambda$, we choose $(T_R,D)$ to give the observed dark matter as shown in \Eq{eq:FIparameters3}.}
\label{fig:FIdilution2}
\end{figure}

We consider the same three values of the coupling $\lambda$  as in \Eq{eq:FIparameters4}, and we choose the dilution factor and the reheat temperature to reproduce the observed DM abundance. There is some degeneracy in the process, and for the purpose of illustration we fix
\be
\lambda = \left\{ \begin{array}{lccll} 
2 \times 10^{-12}  & & & T_R = 1 \, \TeV & D = 1  \\
2 \times 10^{-8}  & & &  T_R = 4.6 \, \GeV & D = 10^8  \\
2 \times 10^{-7}  & & &  T_R = 100 \, \MeV & D = 4.7 \times 10^7  
\end{array} \right.  \ .
\label{eq:FIparameters3}
\ee
The solutions are shown in Fig.~\ref{fig:FIdilution2}. This time we notice a different behavior between the brown and red lines. The reason is a consequence of our choice of $(T_R, D)$: for the (brown) red line the Freeze-In happens during the (non-)adiabatic matter dominated era.

\begin{figure}
\begin{center}
\includegraphics[scale=0.4]{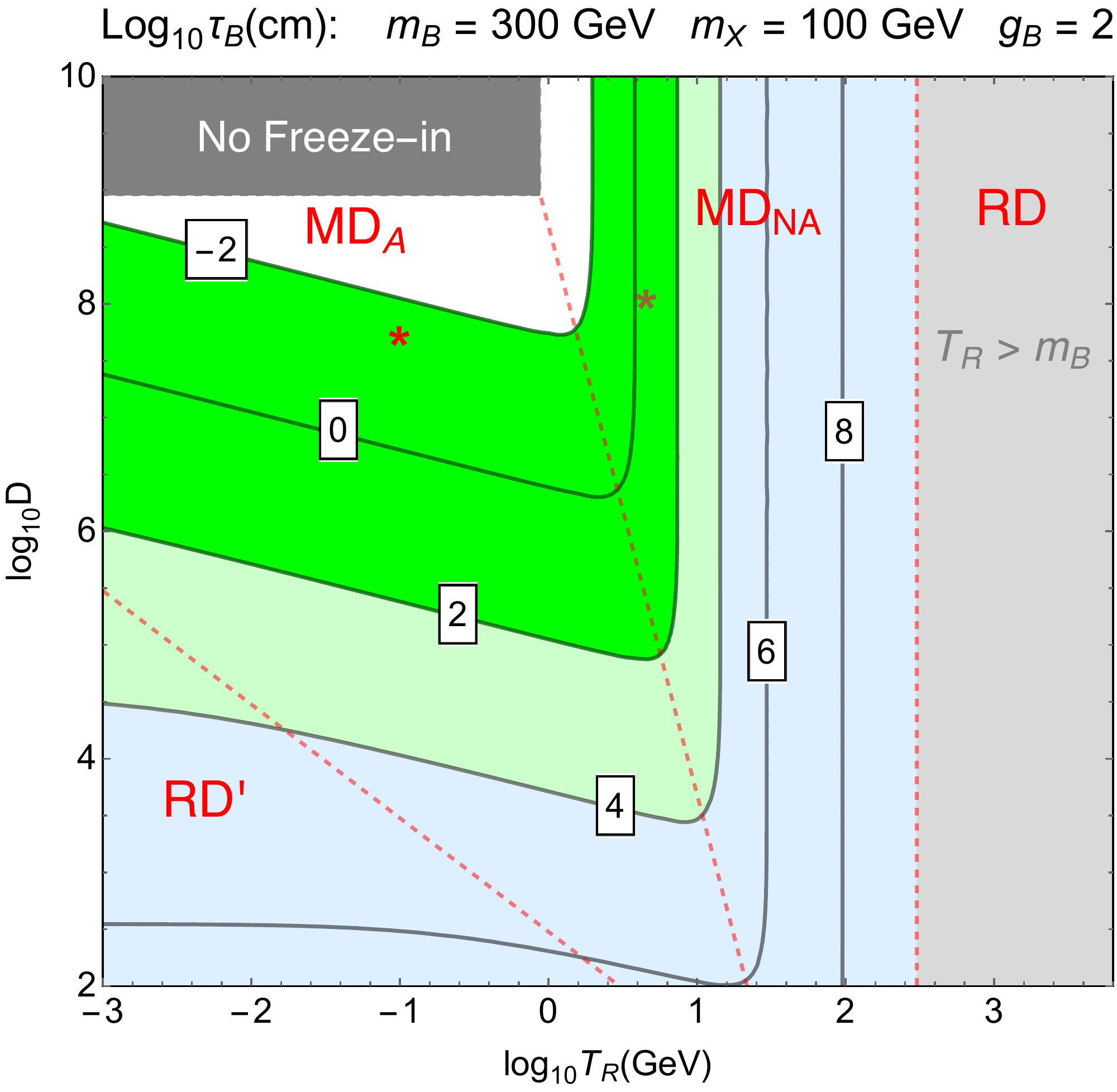} $\quad$ \includegraphics[scale=0.4]{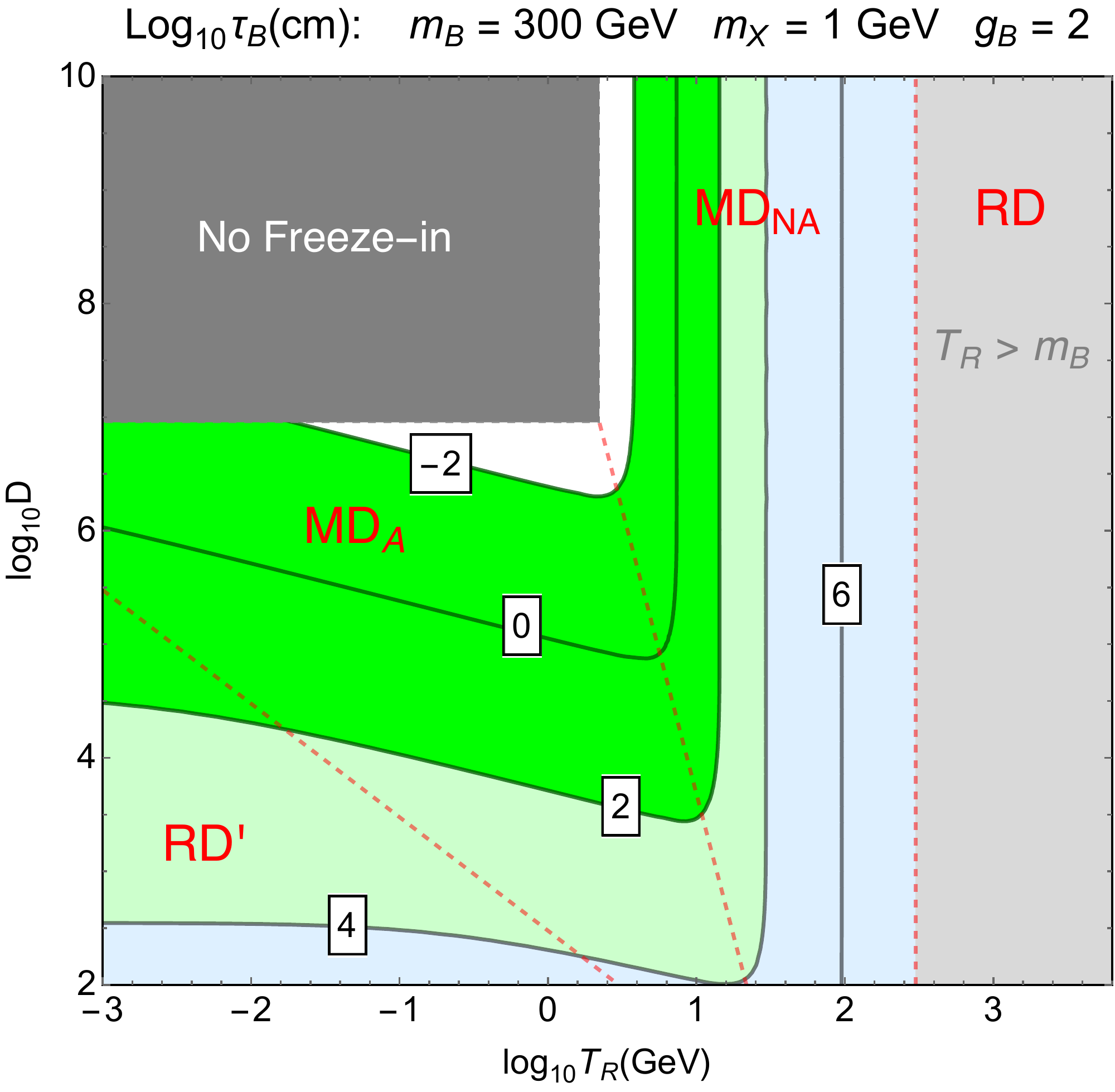} \\ \vspace{0.5cm}
\includegraphics[scale=0.4]{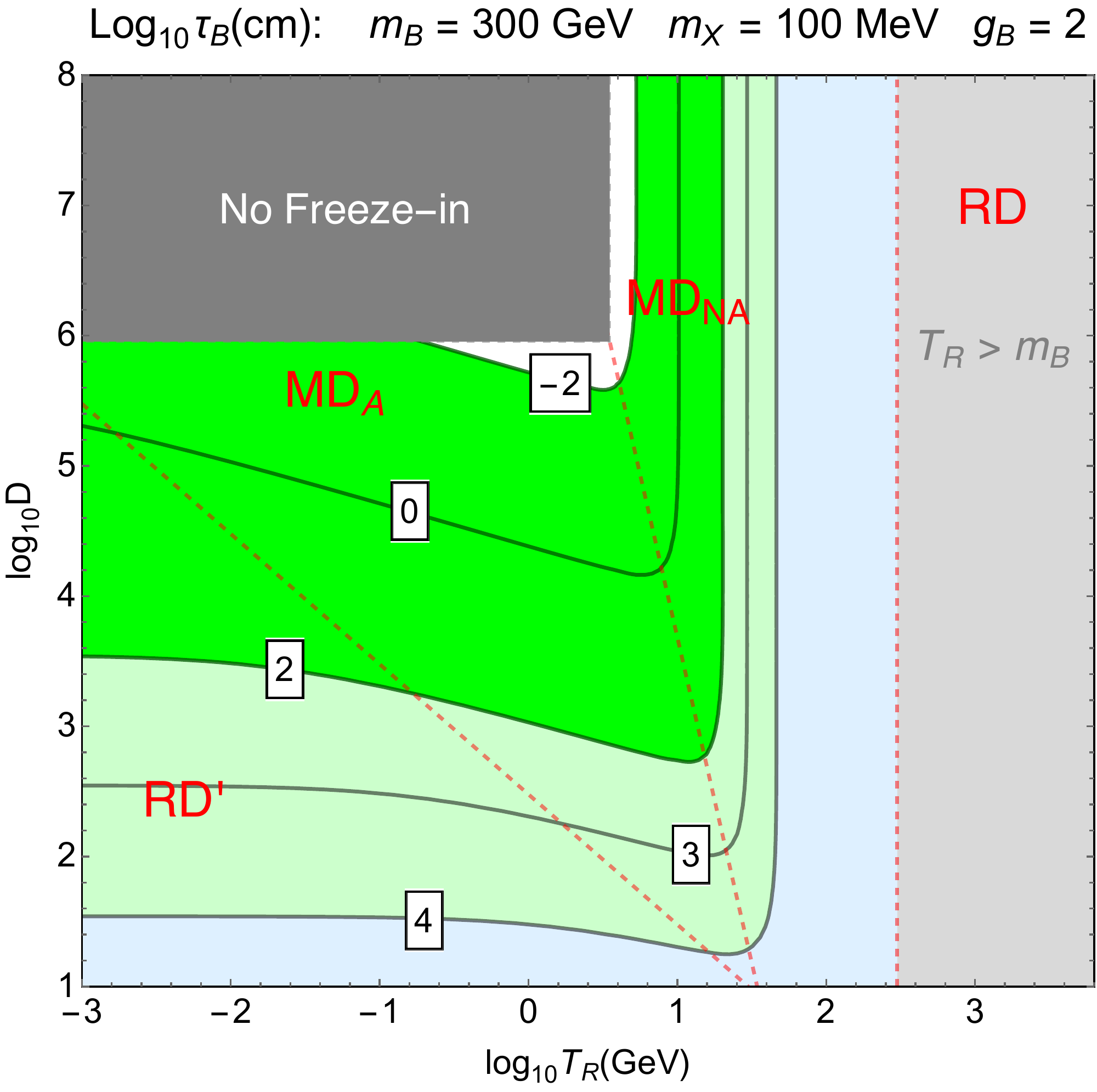} $\quad$ \includegraphics[scale=0.4]{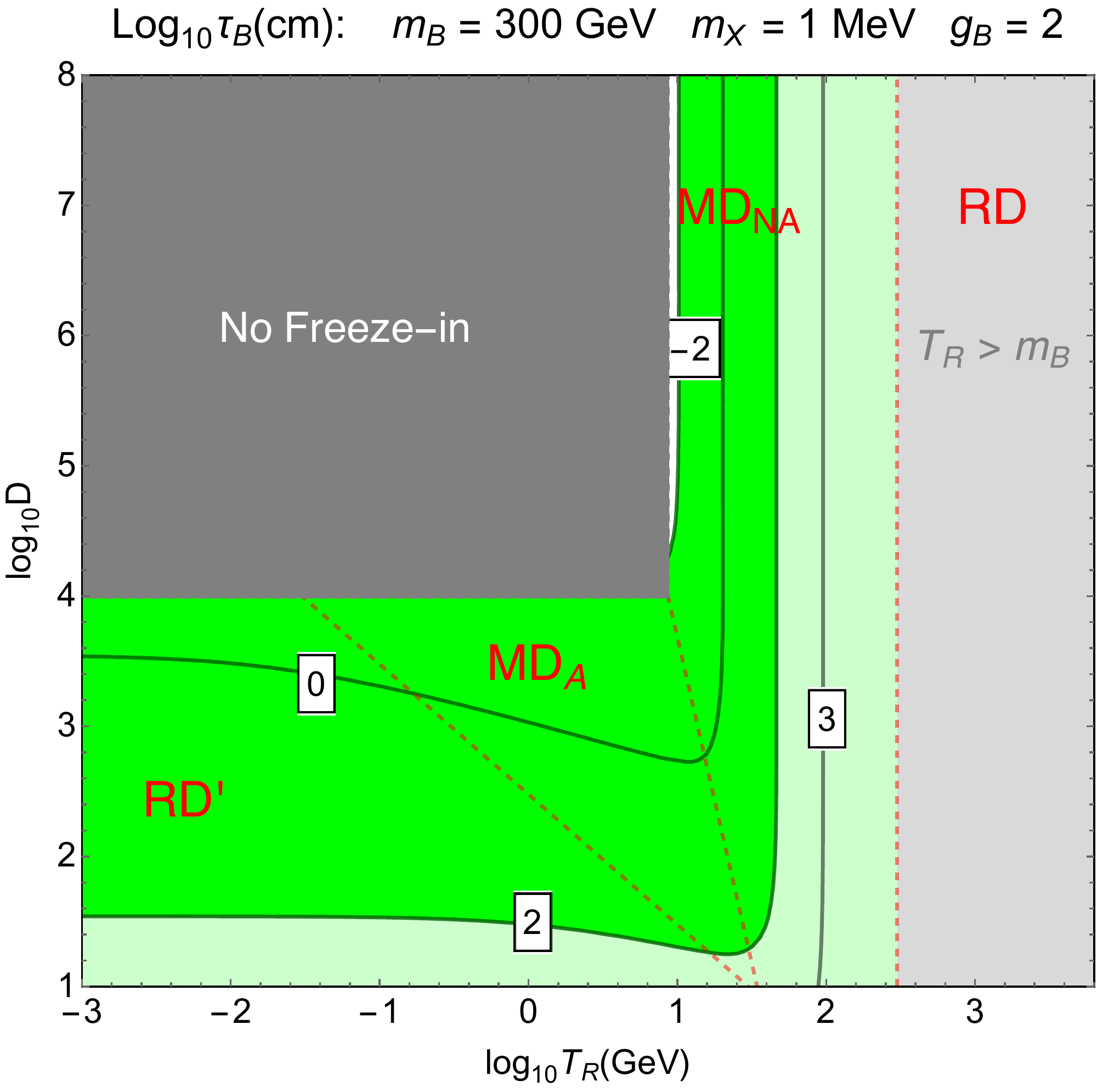}
\end{center}
\caption{Contours of the $B$ lifetime $\tau_B$ (in cm) that give the observed DM abundance via FI in the $(T_R, D)$ plane. We fix $m_B = 300 \, \GeV$ and in each panel we consider different values of $m_X$.  The upper-left gray region does not give FI, as $X$ thermalizes. The three dashed red lines separate four regions where FI occurs during (left to right) RD$'$, MD$_A$, MD$_{NA}$, and RD eras. The red and brown stars in the top-left panel identify the benchmark points chosen in \Eq{eq:FIparameters3} and \Fig{fig:FIdilution2}.  Displaced collider signals occur in the green and blue shaded bands, as described in Table~\ref{table:DisplacedColliderSignals}.}
\label{fig:MasterPlots}
\end{figure}

\begin{figure}
\begin{center}
\includegraphics[scale=0.4]{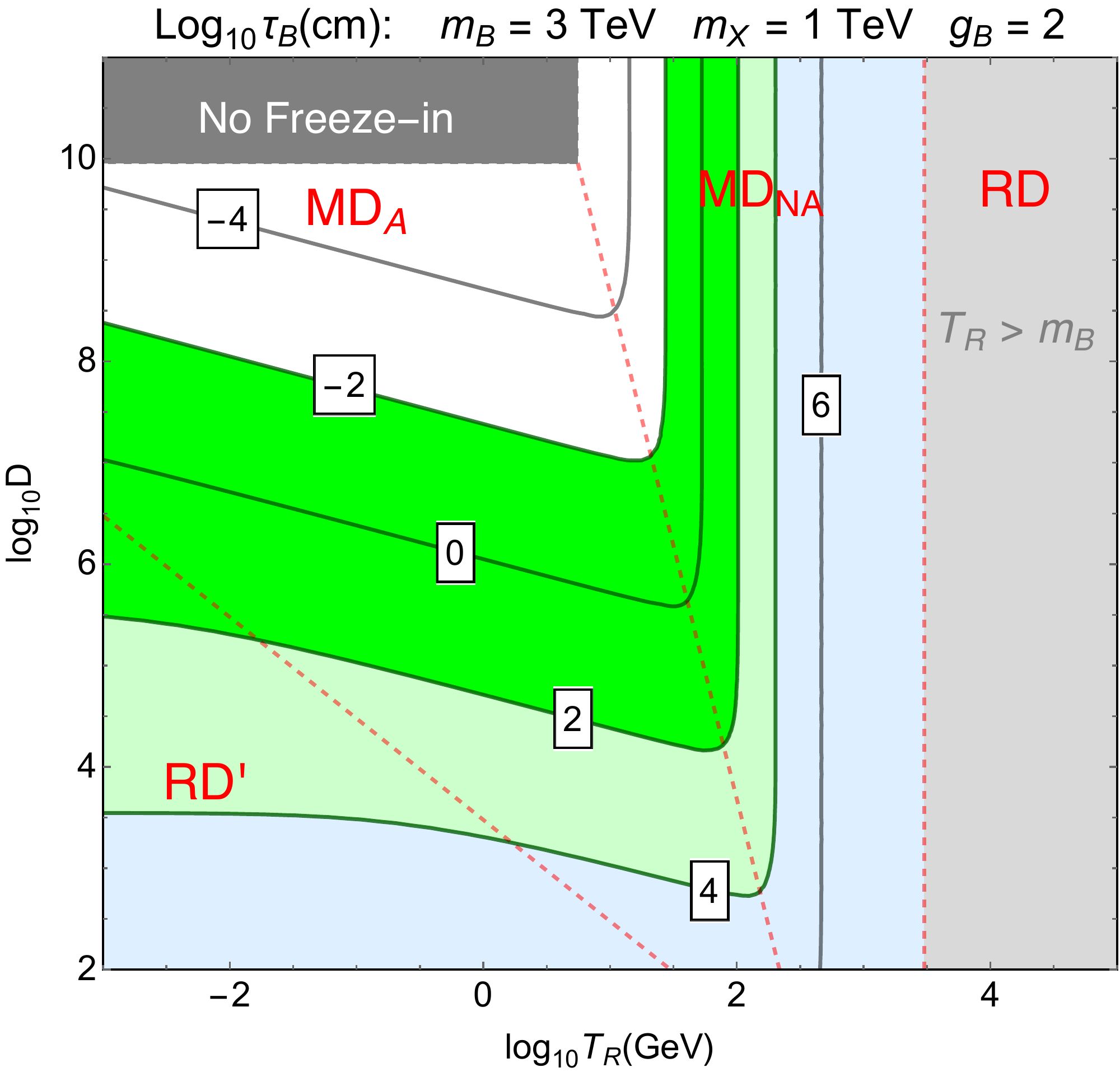} $\quad$ \includegraphics[scale=0.4]{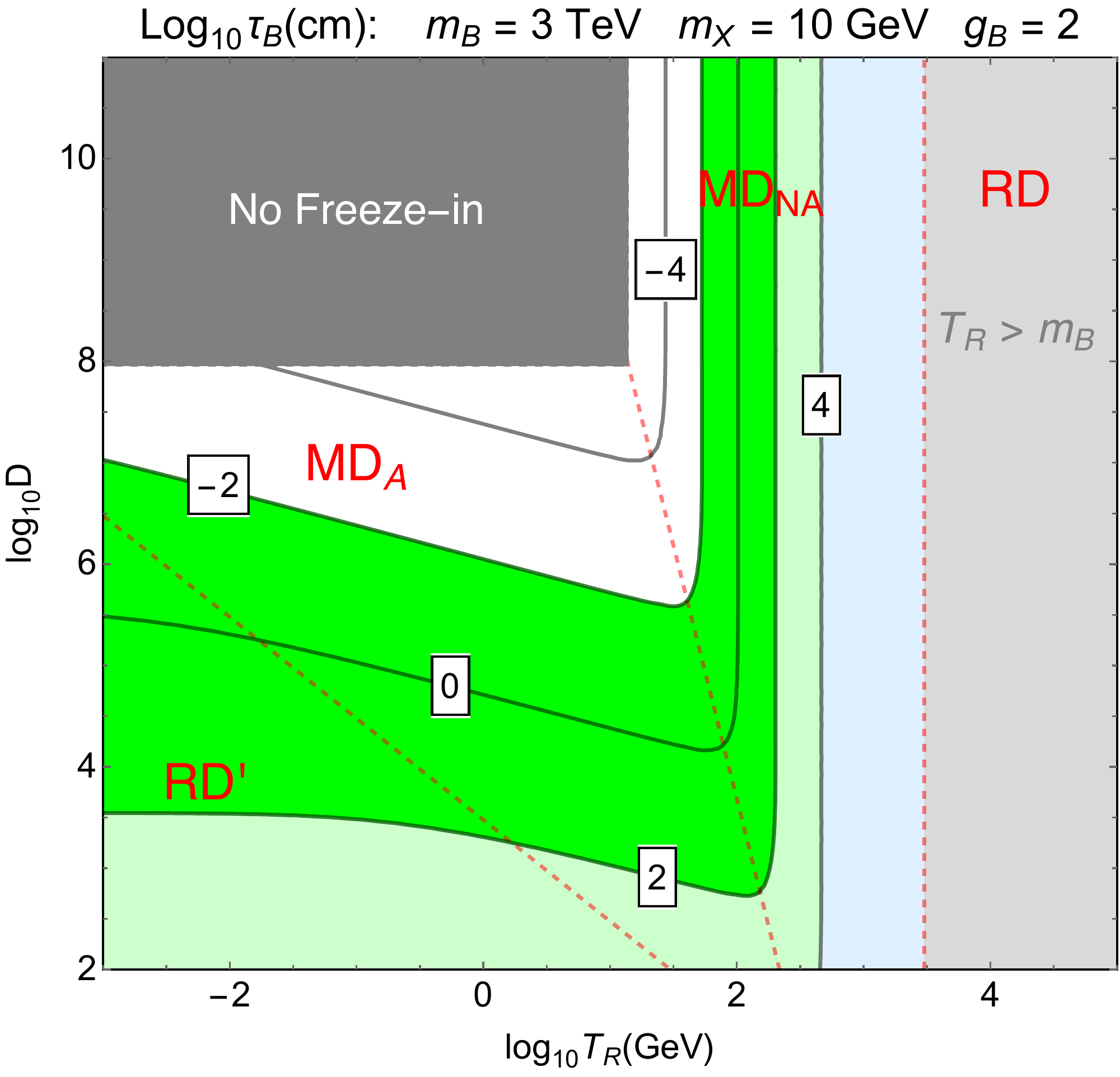} \\ \vspace{0.5cm}
\includegraphics[scale=0.4]{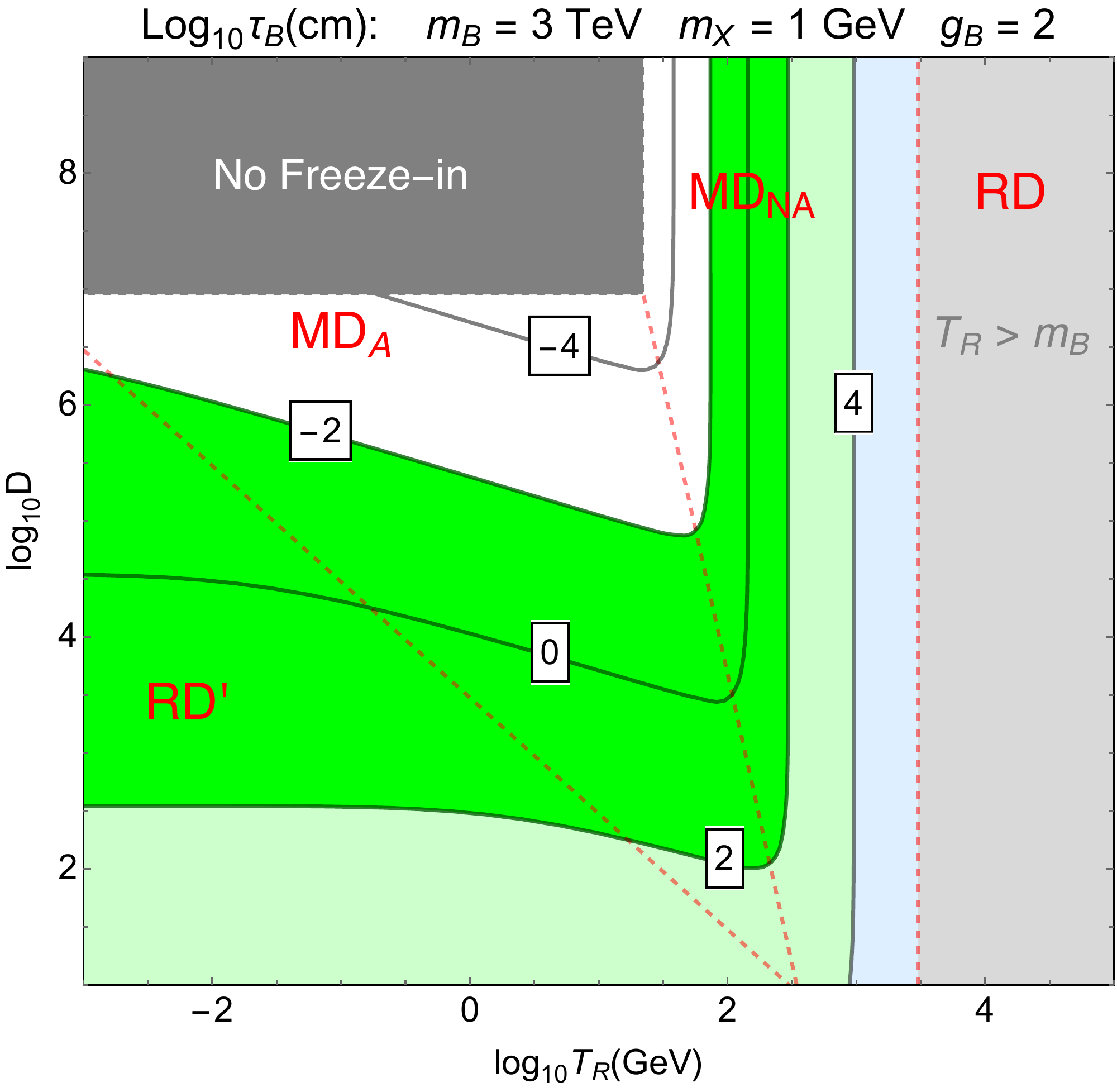} $\quad$ \includegraphics[scale=0.4]{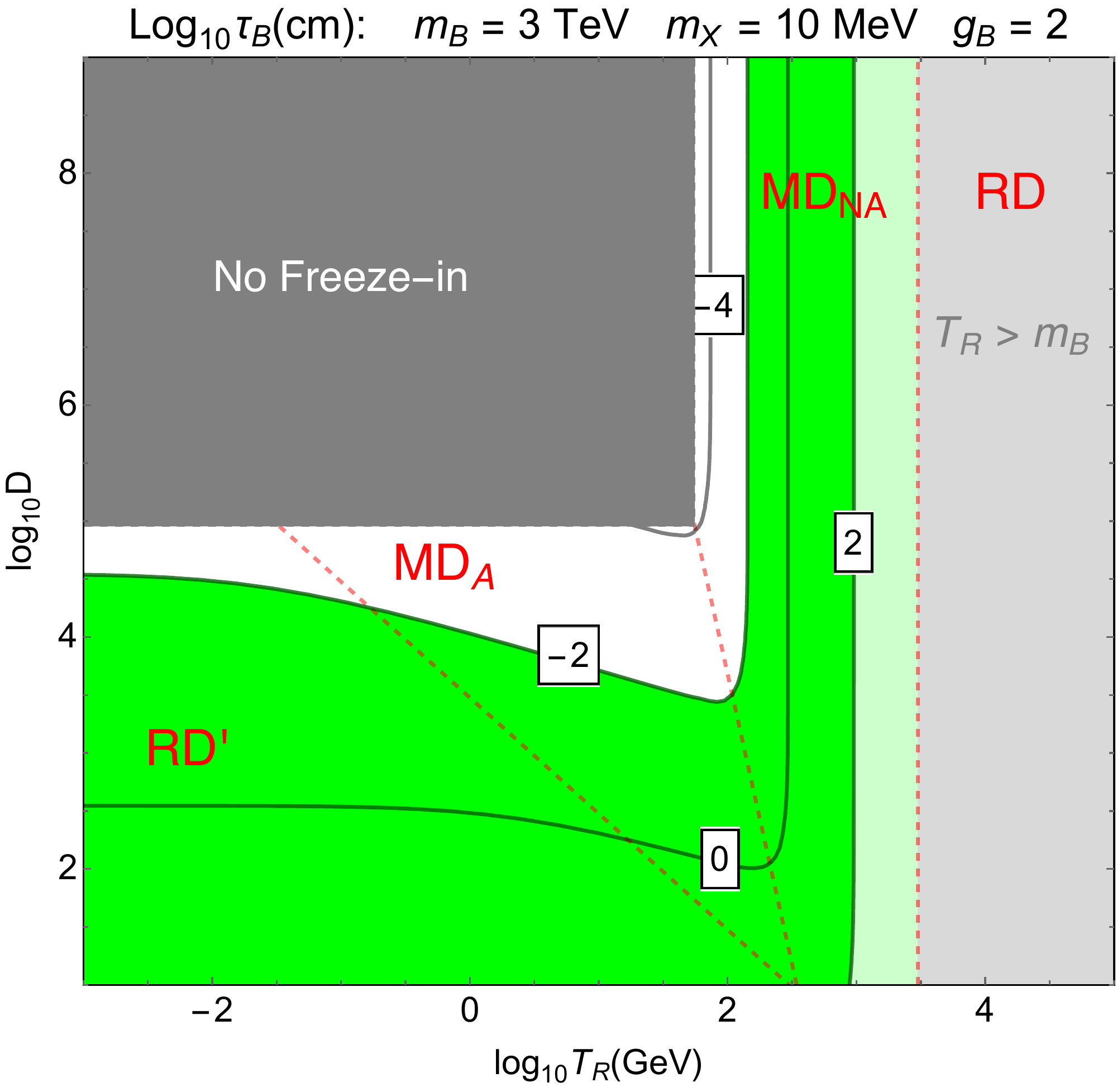}
\end{center}
\caption{Same as \Fig{fig:MasterPlots} but for $m_B = 3 \, \TeV$.}
\label{fig:MasterPlots2}
\end{figure}

Figs.~\ref{fig:MasterPlots} and \ref{fig:MasterPlots2} provide a complete exploration of FI in the $(T_R, D)$ plane, with results shown for different $(m_B,m_X)$. The LOSP mass values of $m_B = 300 \GeV$ (Fig.~\ref{fig:MasterPlots}) and $ 3 \TeV$ (Fig.~\ref{fig:MasterPlots2}) are chosen to be relevant for LHC and a future collider, while values of $m_X$ are chosen to span the very wide range of possible values that is consistent with FI.  Remarkably, the displaced vertex region, shaded in green, is always within the FI domain and covers a substantial part of the parameter space where the observed DM results from FI.

\subsection{Analytic Approximation to Freeze-In Abundance}
\label{subsection:analyticFI}
Many features of the numerical results presented in the previous two sub-sections can be understood from simple analytic approximations.    The FI abundance of $X$ generated during a Hubble doubling temperature at $T$ during era $i$ is
\be
Y^{Prod}_i(T) \, \sim \, \Gamma(T) \; t_i(T) 
\label{eq:YProd}
\ee
where $\Gamma(T)$ is the total rate of $X$ production at temperature $T$.  For clarity numerical factors are dropped in this sub-section, unless they are very large.   The age of the universe at $T$ is
\be
t_i(T)  \sim \frac{1}{H_i(T)}\,\sim  \frac{M_{Pl}}{T^2} \left( 1, \;\;\; \frac{T_R^2}{T^2}, \;\;\; \frac{T^{1/2}}{T_M^{1/2}}, \;\;\;1 \right)
\label{eq:ti}
\ee
with $i=1-4$ running over the (RD, MD$_{NA}$, MD$_A$, RD$'$) eras, with RD the usual RD era at $T<T_R$ and RD$'$ the early one at $T>T_M$.   When FI is induced by an operator of dimension 4, the rate of X production is given by
\be
\Gamma(T) \, \sim \, \lambda^2 m_B
\begin{pmatrix}
\frac{m_B}{T}  \\[0.3em]
\frac{T}{m_B}  
\end{pmatrix}
\label{eq:G4}
\ee
at $T \gsim m_B$.  The upper entry of the vector is for decays, with $m_B/T$ the Lorentz boost factor, while the lower entry is for scatterings. 

This analytic yield $Y^{Prod}_i(T)$ approximates the solution to the Boltmann equations, $Y_X(T)$, for $T \gsim m_B$, where dilution does not have a large effect.  For the case of decays, the power laws in the various eras are $Y^{Prod}_i(T) \propto (1/T^3, 1/T^5, 1/T^{5/2}, 1/T^3)$.  At $x \ll 1$, the RD, MD$_{NA}$ and MD$_A$ slopes are visible in Figs \ref{fig:FIstandard2}, \ref{fig:FIdilution4} and \ref{fig:FIdilution2}, respectively. (In Fig. \ref{fig:FIdilution4} the blue curve starts with MD$_{NA}$ behavior, but the slope decreases as $x$ approaches unity where FI occurs during the RD era.  In Fig. \ref{fig:FIdilution2} the blue curve has $D=1$, so that there is no MD era and the slope reflects RD.)

The contribution to the DM yield today from FI production when the universe had temperature $T$ and was in era $i$ is
\be
Y_{0_i}(T) \, \sim \, \frac{Y^{Prod}_i(T)}{D_i(T)}  
\label{eq:Y01}
\ee
where $D_i(T)$ is the dilution factor from non-adiabatic decays of $M$ during the MD$_{NA}$ era between $T$ and $T_R$
\be
D_i(T) \, \sim \,  \left( 1, \;\;\; \frac{T^5}{T_R^5}, \;\;\; \frac{T_M}{T_R}, \;\;\;\frac{T_M}{T_R} \right).
\label{eq:Di}
\ee
During the MD$_{NA}$ era, \Eq{eq:rhoNA} shows that $n_X \propto T^8$, so that the dilution factor of $Y_X$ scales as $T^5$. As the temperature increases to $T_{NA}$ the dilution reaches $T_{NA}^5/T_R^5$ which by \Eq{eq:TNA} is $T_M/T_R$, the constant dilution factor for production in the MD$_A$ and RD$'$ eras.

Combining \Eq{eq:YProd} --- \Eq{eq:Di} gives
\be
Y_{0_i}(T) \, \sim \,  \frac{\lambda^2 m_B M_{Pl}}{T^2}  
\begin{pmatrix}
\frac{m_B}{T}  \\[0.3em]
\frac{T}{m_B}
\end{pmatrix}
 \left( 1, \;\;\; \frac{T_R^7}{T^7}, \;\;\; \frac{T_R T^{1/2}}{T_M^{3/2}}, \;\;\;\frac{T_R}{T_M} \right).
\label{eq:Y02}
\ee
For decays and scatterings in all eras, FI via interactions of dimension 4 is IR dominated at $T \sim m_B$.  Decays and scatterings then give contributions with the same scaling behavior, giving a total yield today of 
\be
Y_{0_i} \, \sim \,  \frac{\lambda^2 M_{Pl}}{m_B}  
 \left( 1, \;\;\; E \, \frac{T_R^7}{m_B^7}, \;\;\; \frac{T_R m_B^{1/2}}{T_M^{3/2}}, \;\;\;\frac{T_R}{T_M} \right).
\label{eq:Y03}
\ee
Including numerical factors the decay contribution is larger than the scattering one, which is why we have included only decays in our numerical analysis.

For FI during the MD$_{NA}$ era, \Eq{eq:Y02} shows that IR domination is extreme, with a power behavior of $1/T^p$ with $p=8 \, \mbox{or} \, 10$, so that FI production is maximized at temperature $m_B/x_{FI}$ with $x_{FI} \sim 4$.  The high power dependence leads to an abundance from FI on the exponential tale of the Boltzmann distribution for $B$ that is enhanced by a large factor $E$, of order $10^4$, compared to FI at $T\sim m_B$.   Similarly, in Fig \ref{fig:FIdilution4} $Y_X(T)$ peaks at $x >1$.   The $T_R^7$ behavior for the yield during the MD$_{NA}$ era, shown in \Eq{eq:Y03}, leads to the straight lines of slope 7 in Fig. \ref{fig:MasterPlotINF}.

For FI during  the MD$_A$ era,  \Eq{eq:Y03} shows that the yield is proportional to $T_R/T_M^{3/2}$. so that in the MD$_A$ region of Fig. \ref{fig:MasterPlots} the contours of constant $\tau_B$ have $D^3 T_R$ constant.  On the other hand, in the MD$_{NA}$ region of Fig. \ref{fig:MasterPlots} the contours are vertical since they depend only on $T_R$ and not on $T_M$.  The boundary between the MD$_{NA}$ and MD$_A$ eras is given by
\be
D \left( \frac{T_R}{m_B} \right)^5 \, \sim \, \frac{1}{E^{2/3}}.
\label{fig:ANAboundary}
\ee
The boundary between the MD$_A$ and RD$'$ eras is given by $m_B \sim T_M = D T_R$.  

For any $(T_M, T_R)$, knowing which era $i$ dominates FI allows the contours of Fig. \ref{fig:MasterPlots} to be normalized by requiring the observed DM abundance
\be
Y_{0_i} m_X  \, \sim \,  T_{eq}  \simeq 1 \, \mbox{eV},
\label{eq:Teq}
\ee
so that the $B$ decay rate prediction scales as
\be
\Gamma_{B_i} \, \sim \,  \lambda^2_i m_B \, \sim \, \frac{T_{eq}}{M_{Pl}} \, \frac{m_B^2}{m_X}   
 \left( 1, \;\;\;  \frac{m_B^7}{E\, T_R^7}, \;\;\; \frac{T_M^{3/2}}{T_R m_B^{1/2}}, \;\;\;\frac{T_M}{T_R} \right).
\label{eq:GB}
\ee
Including relevant numerical factors, these scaling law predictions can be written as
\begin{eqnarray}
\frac{c \, \tau_B}{\mbox{m.}} &\simeq& 3\times10^6 \left( \frac{300 \GeV}{m_B} \right)^2 \, \left(\frac{m_X}{100 \GeV} \right)  \hspace{1.9in}  (\mbox{RD}) \\
\frac{c \, \tau_B}{\mbox{m.}} &\simeq& 10 \left(\frac{T_R}{10 \GeV}\right)^7
\left( \frac{300 \GeV}{m_B} \right)^9 \, \left(\frac{m_X}{100 \GeV} \right)  \hspace{1.3in}  (\mbox{MD}_{NA}) \\
\frac{c \, \tau_B}{\mbox{m.}} &\simeq& \left(\frac{T_R}{\GeV}\right)\left( \frac{10^5 \GeV}{T_M} \right)^{3/2}
\left( \frac{300 \GeV}{m_B} \right)^{3/2} \, \left(\frac{m_X}{100 \GeV} \right)  \hspace{0.5in}  (\mbox{MD}_A) \\
\frac{c \, \tau_B}{\mbox{m.}} &\simeq& \left(\frac{T_R}{10 \MeV}\right)\left( \frac{30 \GeV}{T_M} \right)
\left( \frac{3 \TeV}{m_B} \right)^2 \, \left(\frac{m_X}{10 \GeV} \right)  \hspace{1in}  (\mbox{RD}')
\label{eq:tauB}
\end{eqnarray}
leading to the behaviors of Figs  \ref{fig:MasterPlotINF}, \ref{fig:MasterPlots} and \ref{fig:MasterPlots2}.
The value of the coupling $\lambda_i$ required for dark matter genesis during era $i$ can be estimated from (\ref{eq:GB}).

For FI during era $i$ to give the observed DM abundance, the peak abundance reached before dilution is
\be
Y^{Peak}_i  \, \sim \, \frac{T_{eq}}{m_X}     \, D_i
\label{eq:Ypeak}
\ee
with $D_i$ evaluated at $T \sim m_B/x_{FI}$ for the MD$_{NA}$ era.  This result for $Y^{Peak}_i$ breaks down for MD$_{NA}$, MD$_A$, RD$'$ when it gives a result larger than the equilibrium abundance, $Y^{eq}$, so that FI is no longer the appropriate description of the DM production mechanism.  This happens for
\be
\frac{T_M}{T_R} \, \gtrsim  \, Y^{eq} \, \frac{m_X} {T_{eq}} \hspace{0.25in} (\mbox{MD}_A, \mbox{RD}') \hspace{1in} \frac{m_B}{x_{FI}T_R} \, \gtrsim  \, \left( Y^{eq} \, \frac{m_X} {T_{eq}} \right)^{1/5} \hspace{0.25in} (\mbox{MD}_{NA})
\label{eq:NoFI}
\ee
and corresponds to the shaded regions labelled ``No Freeze-in" in Figs. \ref{fig:MasterPlots} and \ref{fig:MasterPlots2}. Substituting these bounds into (\ref{eq:GB}) one discovers that the couplings $\lambda_{i\neq RD}$, while much larger than $\lambda_{RD}$, are small.

\subsection{Freeze-In from higher dimensional operators}
\label{subsec:FIn>4}

As the dimension $d$ of the interaction inducing $B \rightarrow X$ increases, so the scattering cross section for this transition grows more rapidly with temperature.  This has the tendency to convert FI from an IR process dominated by temperatures near $m_B$ to a UV one, dominated at some temperature $T_{UV} \gg m_B$. For example, a $d=5$ operator with coefficient $1/M$ inducing FI during era $i$ leads to a total yield today of
\be
Y_{0_i} \, \sim \,  \frac{M_{Pl} \, T_{UV}}{M^2}  
 \left( 1, \;\;\;  E_5 \frac{T_R^7}{m_B^6 \, T_{UV}}, \;\;\; \frac{T_R \, T_{UV}^{1/2}}{T_M^{3/2}}, \;\;\;\frac{T_R}{T_M} \right)
\label{eq:Y04}
\ee
replacing the result (\ref{eq:Y03}) for $d=4$. However, in the case of the MD$_{NA}$ era, the IR dominance for $d=4$ is so strong that the scattering remains IR dominated for all $d<8$ \cite{Giudice:2000ex}.   The FI result is obtained from the $d=4$ case by replacing $\lambda$ with $m_B/M$.  The feeble strength of the interaction necessary for dark matter to originate from FI can be understood as arising from a high mass scale $M \gg m_B$ of size
\be
M^2 \, \sim \, \frac{M_{Pl}}{T_{eq}} \, \frac{T_R^7 \, m_X}{m_B^6}.
\label{eq:M}
\ee
The resulting lifetime, $\tau_B = \tau_B(T_R, m_X, m_B)$, has the same form as for $d=4$ shown in Fig. \ref{fig:MasterPlotINF}, except the normalization is changed because FI may be dominated by scattering rather than decays and as $d$ increases the enhancement factors $E_d$ decrease.

\section{Supersymmetric Theories with Axino Dark Matter}
\label{section:axino}
 The QCD axion is the most elegant solution to the well known CP problem of strong interactions. If supersymmetry is relevant to our universe the details of how the axion solution is implemented can play a crucial role in electroweak physics. This happens if the $H_U H_D$ operator is charged under the Peccei-Quinn (PQ) symmetry whose breaking delivers the QCD axion. If this is the case the usual supersymmetric $\mu$-term is not allowed in the superpotential but must be generated after PQ breaking by operators of the form~\cite{Kim:1983dt}
 \be
 W\supset\frac{\mathcal O^{(n)} H_U H_D}{M_*^{n-1}}
 \label{eq:Wdef}
 \ee
where $\mathcal O^{(n)}$ is a dimension $n$ operator which we assume to be the product of SM gauge singlet superfields with the right PQ charges.  The effective $\mu$-term comes from the vev of $\mathcal O^{(n)}$
\be
\mu=\frac{\langle\mathcal O^{(n)}\rangle}{M_*^{n-1}}.
\ee

The fields that break the PQ symmetry have PQ charges $q_i$ and vevs $V_i$ and $V^2=\sum_i q_i^2 V_i^2$ is the effective PQ breaking scale.
 The axion superfield $A$
 \be
 A=\frac{s+ia}{\sqrt 2}+\sqrt 2\theta\, \tilde a+\theta^2 F
 \label{eq:axionsupermultiplet}
 \ee
transforms under a PQ transformation with real parameter $\alpha$ as $A\to A+i\alpha V$. The axion supersymmetric effective field theory just below $V$ has self interactions and interactions with the Higgs supermultiplets given by \cite{Bellazzini:2011et}
\be
{\cal L}(A, H_{U,D}) \, = \, \int d^4\theta \sum_i V_i^2  e^{q_i(A+A^\dagger)/V}+\mu\int d^2\theta\,  e^{q_{\mathcal O} A/V} H_U H_D+{\textrm{h.c.}}
\ee
where $q_{\mathcal O}$ is the PQ charge of the operator $\mathcal O^{(n)}$. Expanding this Lagrangian gives
\be\label{axinolag}
\int d^4 \theta\, \left(A A^\dagger+\frac{b}{2V} A^2A^\dagger+{\textrm{h.c.}}+\ldots\right)+\int d^2\theta\, \mu H_U H_D\left(1+\frac{q_{\mathcal O}A}{V}+\frac{q^2_{\mathcal O}A^2}{2V^2}+\ldots\right)+{\textrm{h.c.}}
\ee
where  $b=\sum_i q_i^3V_i^2/V^2 $. Additional terms in the Lagrangian give interactions of the axion supermultiplet with supersymmetric gauge field strengths $W_a$
\be\label{anomaly}
{\cal L}(A, W_a) \, = \, \int d^2\theta \sum_a-c_a \, \frac{g_a^2}{16\pi^2 V} \, A\, W^{a\alpha} W^a_\alpha.
\ee
To compute axino FI we work in a basis where the coefficients $c_a$ get contributions only from integrating out matter, charged under both the SM gauge group and PQ symmetry, that has a mass larger than the scale $V$.

\subsection{Axino production in standard cosmology}

In the limit of of unbroken SUSY, the supermultiplet in \Eq{eq:axionsupermultiplet} contains degenerate particles, that are massless up to a small non-pertubative QCD contribution. The axino $\tilde a$ acquires a highly model dependent mass after SUSY breaking. Given its very weak interactions with visible fields, the axino could easily have dramatic consequences for cosmology regardless of where it sits in the spectrum of superpartners~\cite{Rajagopal:1990yx}. We take the axino to be the stable Lightest Supersymmetric Particle (LSP), compute its cosmological abundance, and discuss constraints on the theory that allow it to be dark matter. In this sub-section we assume a standard cosmological evolution, i.e. a non-adiabatic matter dominated era following inflation starting at a temperature $T_{\textrm{max}}$ and ending at a temperature $T_R$, followed by radiation domination down to the eV era.

Axino production in the early universe can take place through different mechanisms. Out-of-equilibrium LOSP decays naturally provide cold axino dark matter~\cite{Covi:1999ty}. Scattering and decays of particles in the primordial plasma give additional contributions to the axino relic density. This can be due to processes mediated by the super-gauge axino-gluino-gluon interaction~\cite{Covi:2001nw} or the axino-higgsino-Higgs Yukawa coupling in DFS models~\cite{Chun:2011zd}. 

Assuming $T_R>\mu$, Eqs.~(\ref{axinolag}) and (\ref{eq:RDFIresult}) allow us to calculate the axino FI yield coming from higgsino decay $\tilde h\to h \tilde a$. In particular the relevant dimension 4 operator in (\ref{axinolag}) gives $\lambda = (q_{\mathcal O}/2)(\mu/V)$.  With $\mu\gg m_h, m_{\tilde a}$ and negligible mixing with the electroweak gauginos, the decay widths of the charged and neutral Dirac components are equal and given by
\be
\Gamma( \tilde h^{\pm}\to W^{\pm} \tilde a)=\Gamma( \tilde h^{0}\to Z\tilde a, \, h\tilde a)\approx \frac{q^2_{\mathcal O}\mu^3}{32\pi V^2}.
\ee
The decay widths do not depend on $\tan\beta$ under the assumption of unmixed higgsinos. These widths define the higgsino lifetime by $\tau=\Gamma^{-1}$. Notice, furthermore, that if the B particle is the higgsino $g_B=8$ in Eq.~(\ref{eq:RDFIresult}), giving
\be\label{axinoFIRD}
Y_{\tilde a}(\mbox{FI}) \, \approx \, 9\times 10^{-4}\,q^2_{\mathcal O}\left(\frac{106.75}{g_*}\right)^{3/2}
\left(\frac{\mu}{300 \, \GeV}\right) \left(\frac{10^{10}\,\GeV}{V}\right)^2.
\ee

An additional contribution to the yield of $\tilde a$ arises if some of the coefficient $c_a$, say $c_3$, are non-vanishing. This contribution is UV dominated and arises from processes of the kind $GG\to G\tilde a$ where $G$ is either a gluon or a gluino and has been calculated in \cite{Strumia:2010aa} 
\be
Y_{\tilde a}(\mbox{UV}) \, \approx \, 1.7\times 10^{-3}\, c_3^2 \left(\frac{2\times106.75}{g_*}\right)^{3/2}\left(\frac{T_R}{10^5 \, \GeV}\right)\left(\frac{10^{10}\,\GeV}{V}\right)^2.
\label{axinoUV}
\ee
This growth with $T_R$ stops once $T_R$ exceeds the mass of the heaviest fermion carrying both SM and PQ charges \cite{Bae:2011jb}, so that this UV contribution is highly model dependent.  In contrast, the FI yield (\ref{axinoFIRD}) is generic to all DFS theories.  The axino abundance is given by the FI mechanism only if $Y_{\tilde a}(\mbox{FI})  \, > \, Y_{\tilde a}(\mbox{UV})$ and $Y_{\tilde a}(\mbox{FI})  \, < \, Y_{\tilde a}^{\textrm{eq}}\approx 2\times 10^{-3}$, the thermal abundance.\footnote{An additional UV dominated contribution to the axino abundance comes from dimension 5 operator of the form $\mu/V^2\,\tilde a\tilde a H_U H_D$. These contribution are suppressed by a factor $\mu^2/V^2$ so that they are irrelevant for weak scale supersymmetry for the values of $T_R$ we are going to consider.}

Defining the axion decay constant, $f$, by the interaction with the gluon field strength $G$ in the low energy effective theory, $(g_3^2/32 \pi^2) (a/f) G \tilde G$, gives
\be
\frac{1}{f} = \frac{c_3+c_3^{IR}+3q_{\mathcal O}}{\sqrt 2\, V}\,,
\label{eq:f}
\ee
where $c_3^{IR}$ is the contribution to the anomaly from non SM fields whose masses are below the scale $f$ and the last term in the sum accounts for the anomalous contribution coming from the SM quarks.  For the choice of parameters made in Eqs.~(\ref{axinoFIRD}) and (\ref{axinoUV}), the axino yield is very high. The observed dark matter density will be exceeded unless the axino mass is far below the supersymmetry breaking scale and/or $f$ is very large.  This is especially true if $T_R \gsim 2.10^5 \GeV/c_3^2$, when the yield is dominated by UV scattering.   Large $f$ and/or small $m_{\tilde a}$ are certainly possible, and are the conventional approach to this axino over-abundance problem, but they place severe restrictions on the theory.   

The simplest theories giving the superpotential as in \Eq{eq:Wdef} have $\mathcal O^{(n)}$ with dimension $n=2$, giving $f^2 < \mu M_{Pl}$ for $M_* < M_{Pl}$, so that the yields of Eqs.~(\ref{axinoFIRD}) and (\ref{axinoUV}) are indeed very large.  For theories with larger $n$, a large value of $f$ yields too much axion dark matter unless the initial axion misalignment angle is adequately tuned to small values. The tree-level axino mass is bounded from above by $m_{\tilde{a}} \lesssim \widetilde{m}^2 / f$~\cite{Tamvakis:1982mw}, with $\widetilde{m} \sim \mu$ the scale of SUSY breaking. This limit applies even taking into account the soft SUSY breaking terms~\cite{Rajagopal:1990yx}. This mass would be small enough to avoid any overabundance. However, this tree-level prediction can easily be spoiled. It is quite generic for the axino to pick up a mass of order the gravitino mass via higher dimensional operators coupling the axion supermultiplet to SUSY breaking fields~\cite{Cheung:2011mg}. If SUSY is broken at a high scale, as in gravity mediation, the axino will generically pick up a weak scale mass. Even for lower mediation scales there can be loop-mediated SUSY breaking contributions to the axino mass (first pointed out in Ref.~\cite{Moxhay:1984am} within the context of supergravity theories) depending on the UV completion of Eq.~(\ref{axinolag}).

In this section we pursue an alternative solution to this axino-overproduction problem: we take $T_R$ below the higgsino mass so that FI occurs during the MD era.  The FI yield (\ref{axinoFIRD}) is then suppressed by a combination of less available time for production and subsequent dilution from entropy production, as in (\ref{eq:Y03}).  From the above discussion we assume the axino mass is not far below the weak scale
\be\label{axinorange}
10^{-3}\lesssim \frac{m_{\tilde a}}{\mu}\lesssim 1.
\ee
The UV contribution (\ref{axinoUV}) is similarly suppressed and, as discussed in the next sub-section, we will be interested in situations where it is sub-dominant. We find that axino dark matter then predicts a restricted range of $f$ and that in much of the parameter space the reactions $\tilde h\to h \tilde a, \, Z \tilde a$ and $\tilde h^\pm \rightarrow W^\pm \tilde a$ lead to displaced signals at colliders. 



\subsection{Axino Freeze-In during an early matter dominated era}
Let us review the assumptions which will go into this section. We assume the strong CP problem to be solved by a supersymmetric DFS axion.  The axino mass is required to be in the range of Eq.~(\ref{axinorange}) and we assume the axino to be the DM. We furthermore assume for simplicity the LOSP to be a higgsino with negligible mixing with the electroweak gauginos.

We begin by requiring there are no fields charged under both the PQ and SM gauge symmetries that are heavier than the TeV scale, so that $c_a, c_a^{IR}=0$ in Eq.~(\ref{anomaly}) and there is no UV scattering yield from temperatures above the TeV scale. In this case, after the rescaling of Eq.~(\ref{eq:f}), the higgsino decay widths depend only on $f$,
\be
\Gamma( \tilde h^{\pm}\to W^{\pm} \tilde a)=\Gamma( \tilde h^{0}\to Z\tilde a, \, h\tilde a)\approx \frac{\mu^3}{144\pi f^2}.
\label{eq:axinodecay}
\ee

The general analysis of FI in a cosmology with an intermediate MD era performed in Sec.~(\ref{sec:FI}) can then be directly translated to this model using Eq.~(\ref{eq:axinodecay}), for both the numerical and analytic analyses. For any point of the general parameter space depicted in Fig.~(\ref{fig:Histories}) we are thus able to determine the value of $V$, and hence $f$, that gives the observed dark matter abundance.

\begin{figure}
\begin{center}
\includegraphics[scale=0.37]{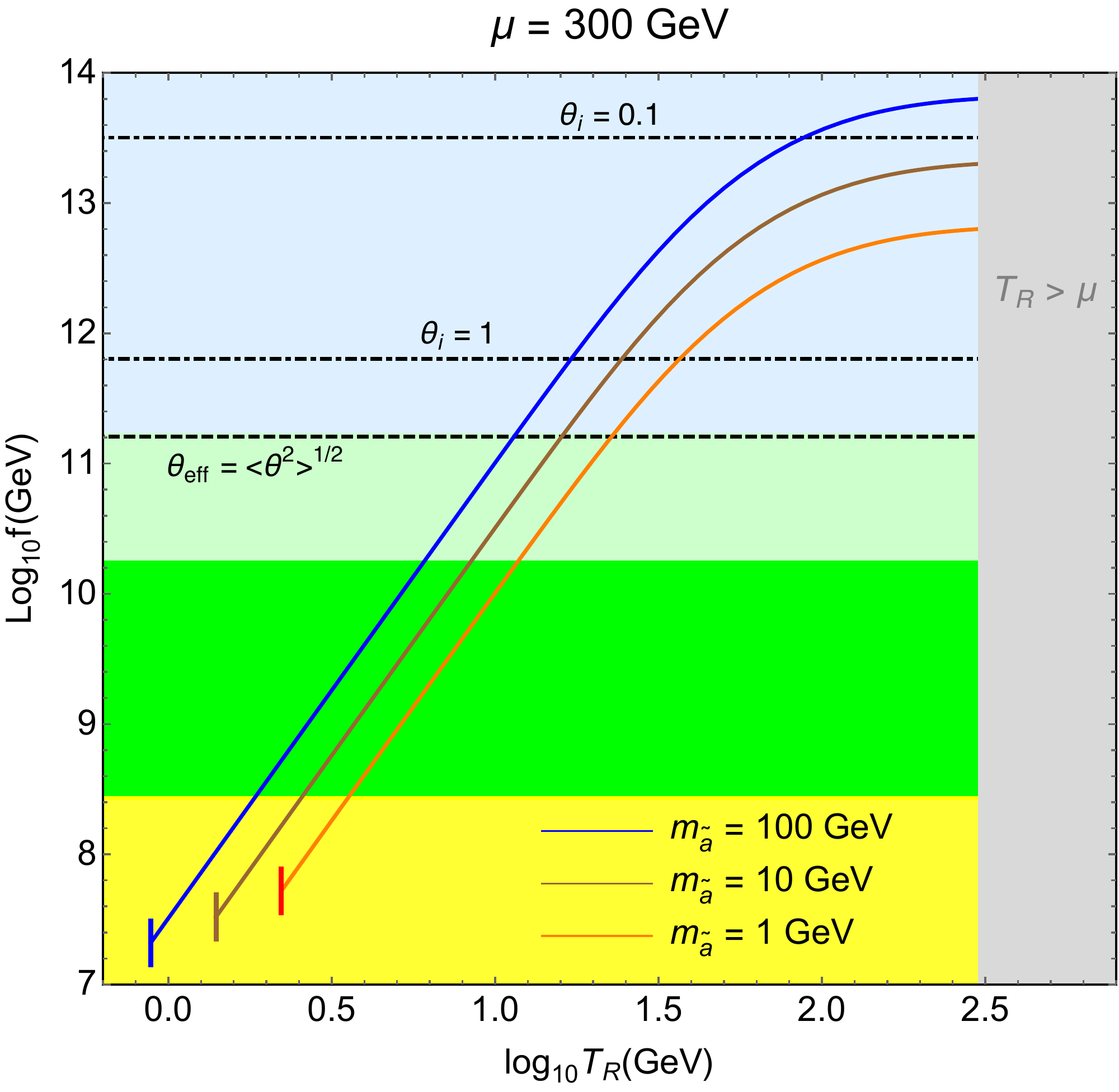} $\quad$
\includegraphics[scale=0.37]{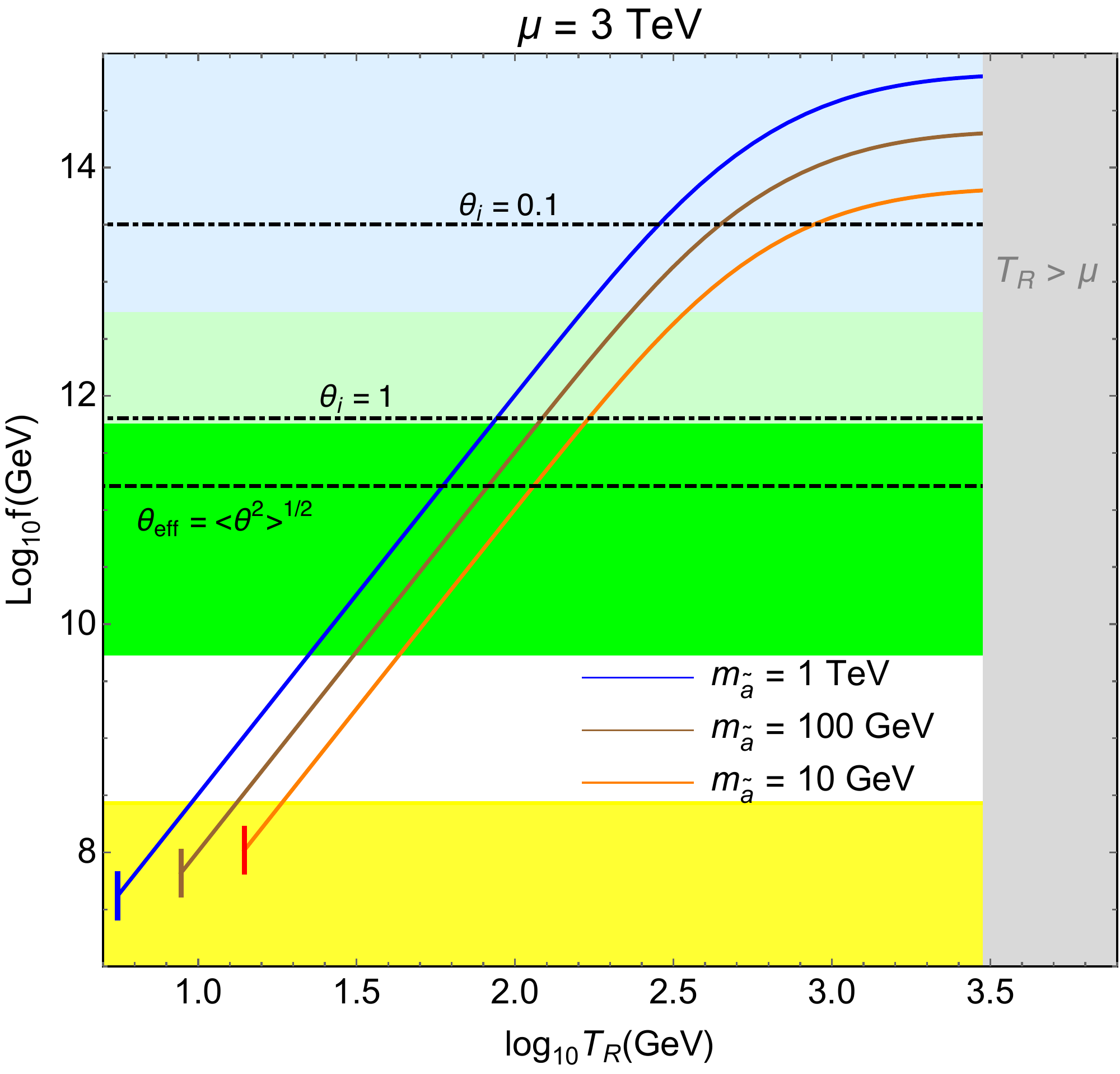} 
\end{center}
\caption{$f$ as a function of $T_R$ that yields axino dark matter for $\mu = 300 \, \GeV$ (left) and $\mu = 3 \, \TeV$ (right). In each case we show results for three different axino masses. Above the black dashed line the axion abundance is above the observed value in the post-inflationary case, while the dot-dashed lines refer to the pre-inflationary scenario for various choices of the initial misalignment angle.
The yellow region is excluded by axion emission from white dwarfs. Displaced collider signals occur in the shaded horizontal bands as described in Table~\ref{table:DisplacedColliderSignals}.}
\label{fig:MasterPlotINFAxino}
\end{figure}

\begin{figure}
\begin{center}
\includegraphics[scale=0.4]{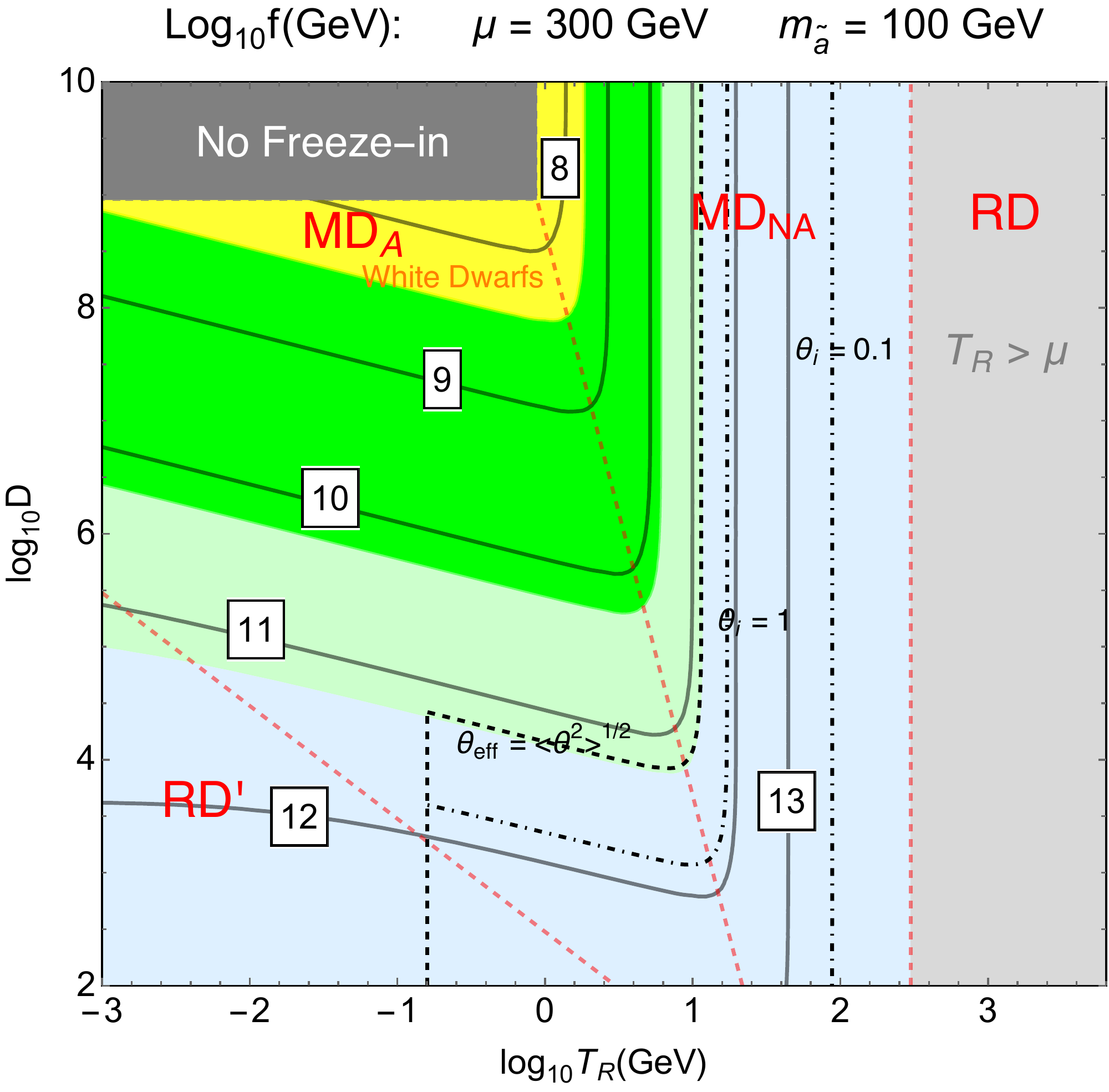} $\quad$ \includegraphics[scale=0.4]{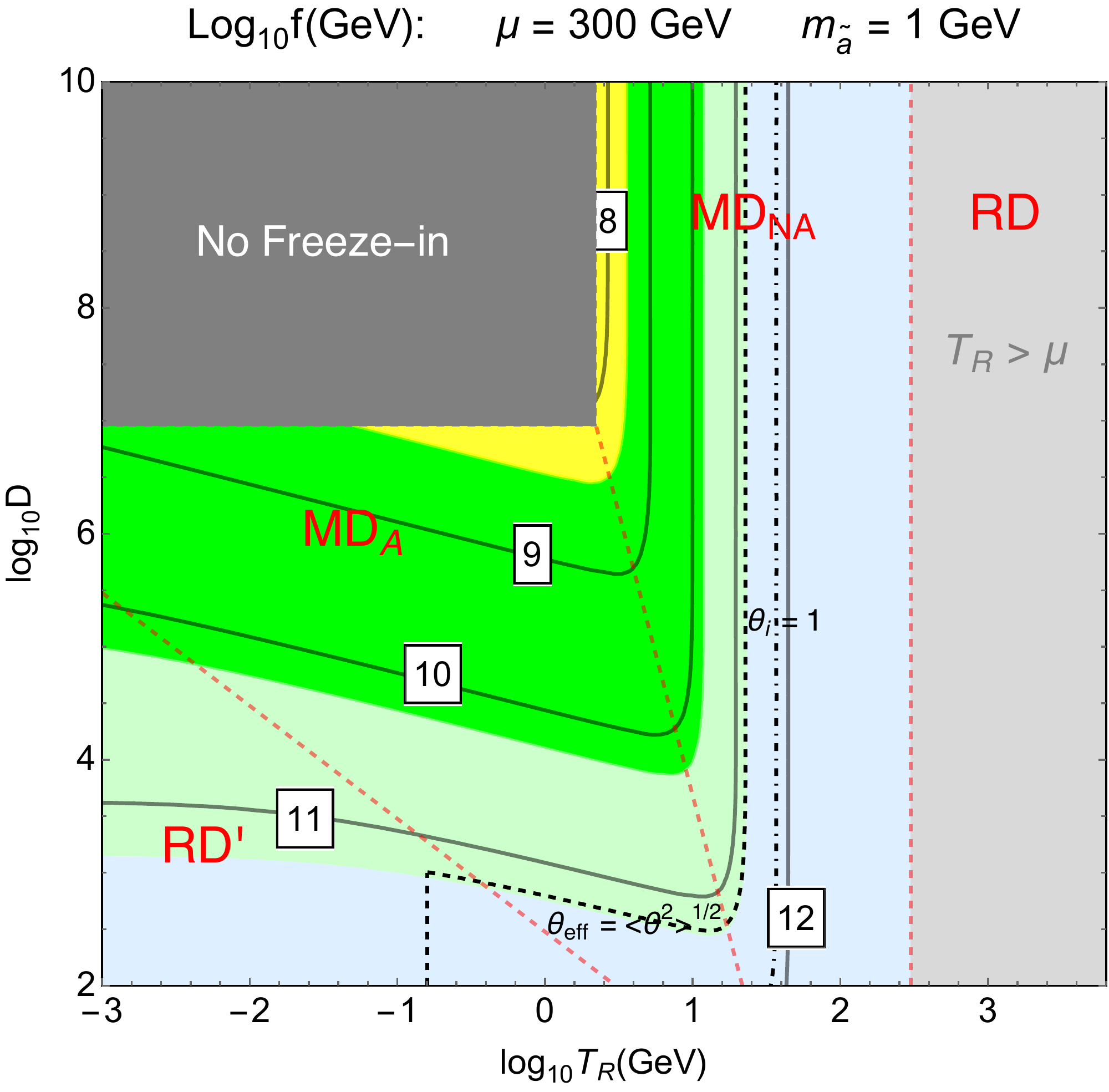} \\ \vspace{0.5cm}
\includegraphics[scale=0.4]{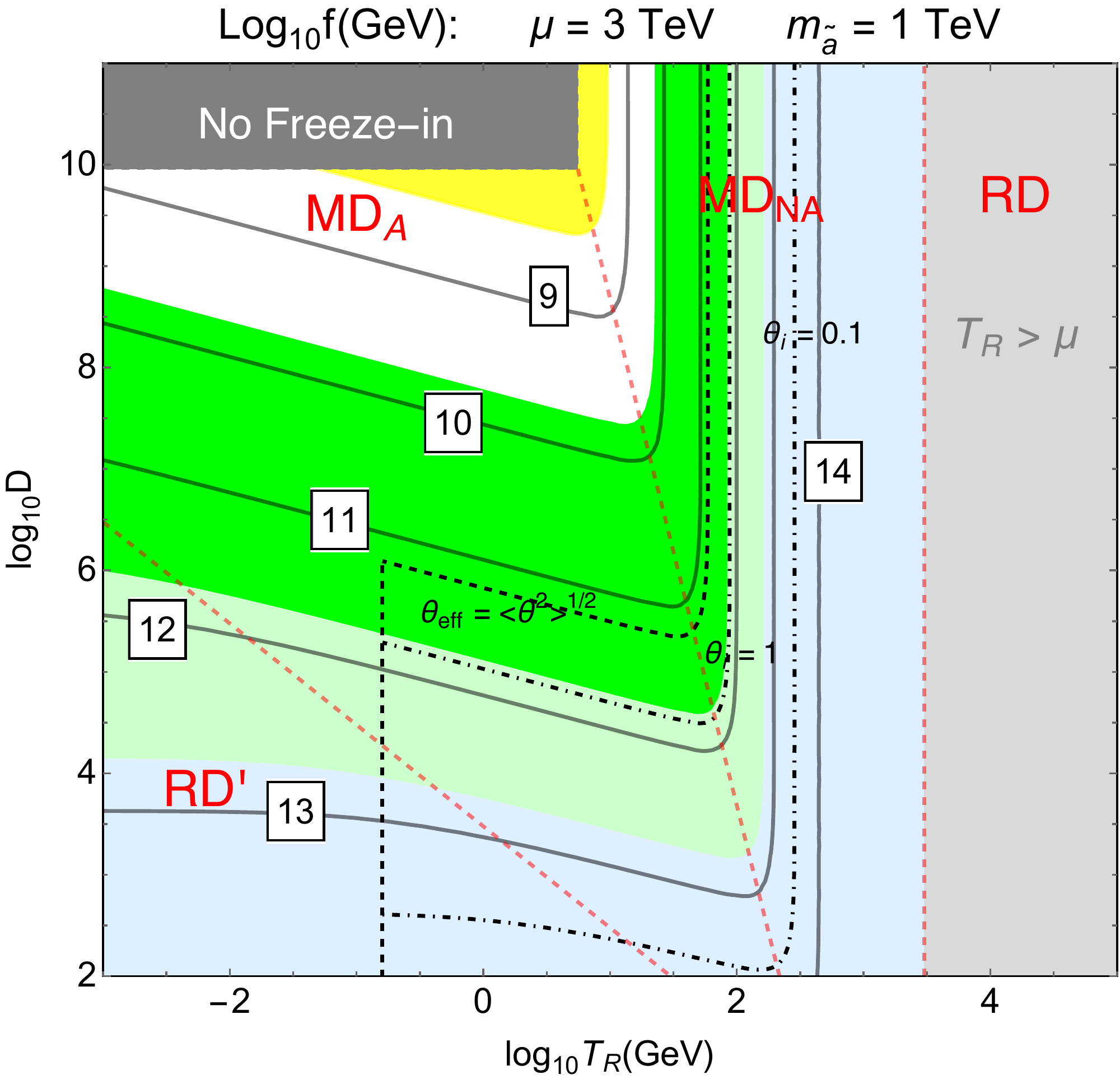} $\quad$ \includegraphics[scale=0.4]{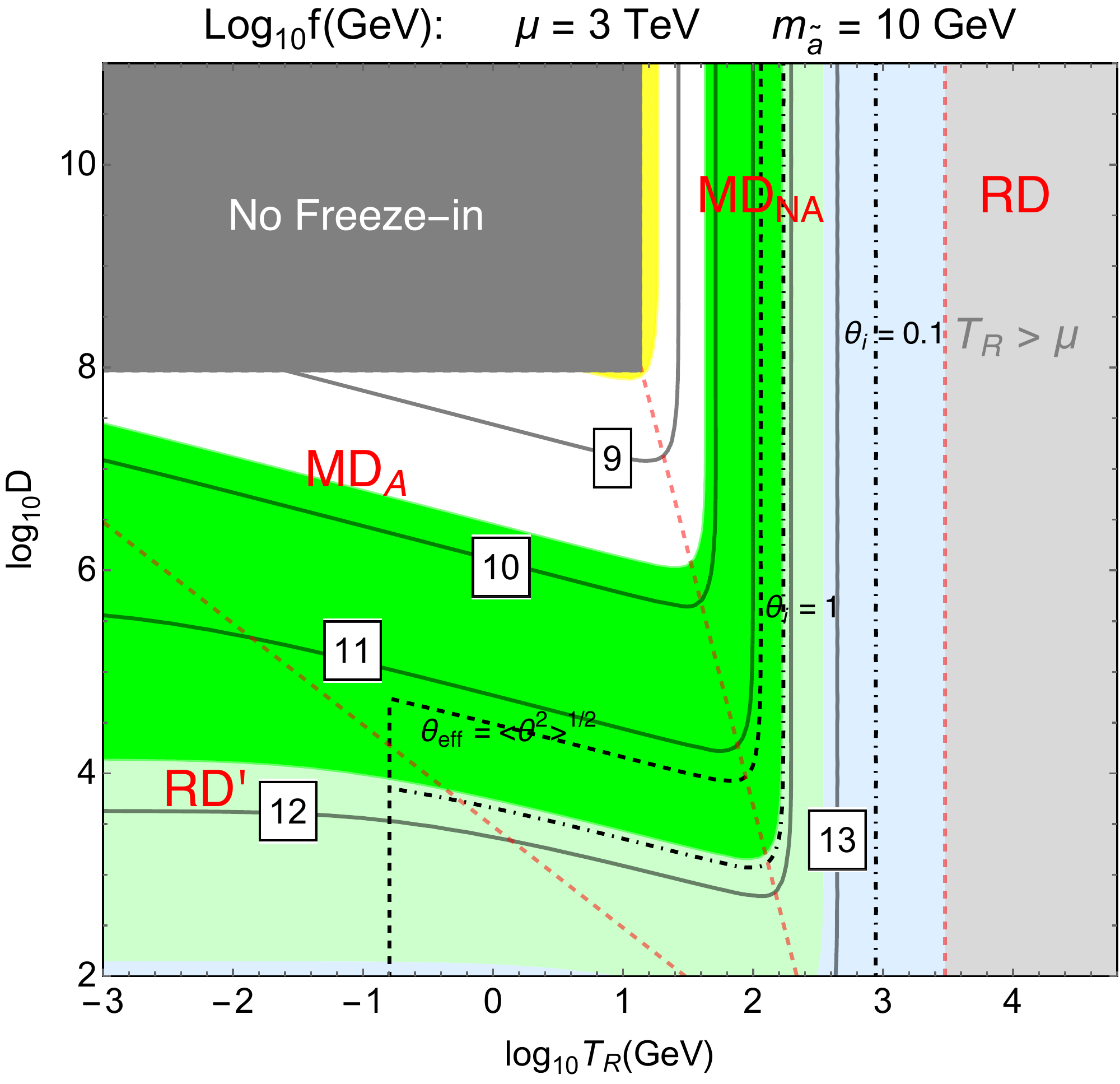}
\end{center}
\caption{Contours of $f$ that give axino DM via FI in the $(T_R, D)$ plane for different values of $\mu$ and $m_{\tilde a}$.  To the right of the black dashed line the axion abundance is above the observed value in the post-inflationary case, while the dot-dashed lines refer to the pre-inflationary scenario for various choices of the initial misalignment angle. The yellow region is excluded by white dwarfs constraints.  Displaced collider signals occur in the green and blue shaded bands, as described in Table~\ref{table:DisplacedColliderSignals}.}
\label{fig:MasterPlotsAxino}
\end{figure}

The numerical results for axino FI are shown in Figs.~(\ref{fig:MasterPlotINFAxino}) and (\ref{fig:MasterPlotsAxino}) for FI during inflationary reheating and during a generic MD cosmology respectively. 
The qualitative behavior of the figures can be understood using the analytical approximations in Sec.~(\ref{subsection:analyticFI}), in particular (\ref{eq:Y03}), and taking $g_B=8$ for the higgsino:
\begin{eqnarray}\nonumber
\frac{f}{\textrm{GeV}}&\simeq& 1.0 \times 10^{11}\left(\frac{T_R}{10\,\GeV}\right)^{\frac{7}{2}}
\left( \frac{300 \GeV}{\mu} \right)^3 \, \left(\frac{m_{\tilde a}}{100 \GeV} \right)^{\frac{1}{2}}  \hspace{1.1in}  (\mbox{MD}_{NA}) \\
\frac{f}{\textrm{GeV}} &\simeq&0.4 \times 10^{11} \left(\frac{T_R}{ \GeV}\right)^{\frac{1}{2}}\left( \frac{10^5 \GeV}{T_M} \right)^{3/4}
\left( \frac{\mu}{300 \GeV} \right)^{3/4} \, \left(\frac{m_{\tilde a}}{100 \GeV} \right)^{\frac{1}{2}}  \hspace{0.2in}  (\mbox{MD}_A) \\\nonumber
\frac{f}{\textrm{GeV}}&\simeq&1.2\times 10^{12}\left(\frac{T_R}{10 \MeV}\right)^{\frac{1}{2}}\left( \frac{30 \GeV}{T_M} \right)^{\frac{1}{2}}
\left( \frac{\mu}{3 \TeV} \right)^{\frac{1}{2}} \, \left(\frac{m_{\tilde a}}{10 \GeV} \right)^{\frac{1}{2}}  \hspace{0.6in}  (\mbox{RD}')
\label{eq:faxion}
\end{eqnarray}

In the region to the right of the black dashed line in Figs.~(\ref{fig:MasterPlotINFAxino}) and (\ref{fig:MasterPlotsAxino}), there is an overabundance of QCD axions produced by the misalignment mechanism. This limit applies to the region of parameter space with a post-inflationary scenario \cite{Hertzberg:2008wr,D'Eramo:2014rna} where the PQ phase transition occurs after inflation. If $T_R$ is above 160\,MeV the entropy generated at the end of the MD era does not dilute the axion density coming from coherent oscillations of the axion field but only the (theoretically uncertain) contribution coming from the decay of topological defects \cite{Davis:1986xc}. For values of $T_R$ below 160\,MeV the misalignment contribution is strongly diluted and negligible. Notice that the red exclusion does not apply for the pre-inflationary axion scenario as one can  tune the initial misalignment angle to arbitrarily small values.

The yellow shaded region in the figures corresponds to the bound coming from axion emission from white dwarfs \cite{Agashe:2014kda}. The bound applies if the axion-electron coupling
\be
\frac{C_e}{2f}\,\partial_\mu a \,\bar e \gamma^\mu\gamma_5 e
\ee
is present, which is typically the case for DFS axions, and is shown for $C_e=1/3$.  The axion energy density bound from misalignment and the white dwarf bound depend on $f$ and are shown in Figs.~(\ref{fig:MasterPlotINFAxino}) and (\ref{fig:MasterPlotsAxino}).

The green shaded regions correspond to higgsino lifetimes that give rise to displaced signals at colliders (with the color code as defined in Tab.~\ref{table:DisplacedColliderSignals}).  The higgsino decays to missing energy and, depending on whether it is charged or neutral, to a $W$ boson or democratically to a Higgs or Z boson. An early study of these signals can be found in Ref.~\cite{Martin:2000eq}, a more recent analysis for LHC collisions was performed in Ref.~\cite{Barenboim:2014kka}. Notice that this signal is distinguishable from the decay of the neutralino NLSP in gauge mediated models where final states containing a photon are possible while they are strongly suppressed in the axino case. 

At a hadron collider, in the most favorable situation, it will be possible to measure both $\mu$ and $m_{\tilde a}$ from kinematical distributions and the higgsino decay width from the displaced vertex. This is enough to extract $T_R$ in the case of FI during post-inflationary cosmology or a corresponding curve in the $(T_R,\, D)$ plane in the case of a generic MD era. In the axino case the decay width is related to the axion decay constant $f$ and this correspondence is 1 to 1 for an unmixed higgsino as in Eq.~(\ref{eq:axinodecay}). Even though the axion is not dark matter in most of the displaced vertex region, low energy experiments looking for axion-mediated long-range forces \cite{Arvanitaki:2014dfa} can measure $f$ independently and provide a crucial verification of the framework discussed in this section.
 
How are these results affected by going to non-minimal DFS models with matter heavier than the TeV scale inducing $c_a, c_a^{IR} \neq 0$ in (\ref{anomaly}) and (\ref{axinoUV})?  For the case of the MD$_{NA}$ cosmology, as occurs after inflation as shown in \Fig{fig:Eras2}, there is no change and \Fig{fig:MasterPlotINFAxino} applies unaltered.  In this case, even though the operator of (\ref{anomaly}) has dimension 5, it induces an axino yield that is IR rather than UV dominated, and this IR contribution is always less than that from higgsino decay.  Furthermore, additional FI yields from heavy colored states are suppressed by the strong IR dominance of MD$_{NA}$.  However, gauge scattering axino production is UV dominated during MD$_A$ and RD$'$ eras, as well as during any RD era that preceded them, and further FI yields from these eras might be significant.  In this case it is model-dependent whether the axino yield is dominated by FI from higgsino decays and \Fig{fig:MasterPlotsAxino} applies only if the FI contribution from higgsinos dominates.

Figs.~(\ref{fig:MasterPlotINFAxino}) and (\ref{fig:MasterPlotsAxino}) demonstrate that, in a wide range of DFS models, the dark matter can be axinos with a mass in the GeV to TeV range, with the dominant production mechanism being FI from higgsino decay during an early MD era.  Over a large fraction of the parameter space not excluded by axion overproduction and white dwarf constraints, there are displace signals at colliders arising from precisely the decays that induced the cosmological abundance, $\tilde h\to h \tilde a, \, Z \tilde a$ and $\tilde h^\pm \rightarrow W^\pm \tilde a$ .  The relevant range of $f$ is low: from the experimental lower bound up to order $10^{12}$ GeV, which can be independently explored by a variety of techniques \cite{Arvanitaki:2014dfa}.

\section*{Acknowledgments}
This work was supported in part by the Director, Office of Science, Office of High Energy and Nuclear Physics, of the US Department of Energy under Contract DE-AC02-05CH11231 and by the National Science Foundation under grants PHY-1002399 and PHY-1316783. F.D. is supported by the Miller Institute for Basic Research in Science. R.C. is supported by the National Science Foundation Graduate Research Fellowship under Grant No. DGE 1106400.

\appendix

\section{Solution of FI Boltzmann Eq. for Any History}
\label{sec:FIsol}

We solve the Freeze-In Boltzmann equation (\ref{eq:BoltzFI}) by employing techniques analogous to the ones used in Refs.~\cite{Chung:1998rq,Giudice:2000ex}. In cases where entropy is not conserved in the evolution, $Y_X$ is not the best variable to use. We define the $X$ comoving number density, $\mathcal{X} = n_X a^3$, and we use the scale factor as time variable $dt= H^{-1}d \ln a$.

The Freeze-In Boltzmann equation in terms of the new variable reads
\be
\frac{d \mathcal{X}}{d \ln a} = \Gamma_B \frac{n_B^{\rm eq}(T(a)) \, a^3}{H(a)}  \,\frac{K_1[m_B/T(a)]}{K_2[m_B/T(a)]}  \ .
\label{eq:FIandDilution}
\ee
The asymptotic comoving density is found by imposing the initial condition $n_X(0) \simeq 0$
\be
\mathcal{X}_\infty = \Gamma_B  \int_0^{\infty} d \ln a^\prime \; 
\frac{n_B^{\rm eq}(T(a^\prime)) \, a^{\prime 3}}{H(a^\prime)}  \,\frac{K_1[m_B/T(a^\prime)]}{K_2[m_B/T(a^\prime)]}  \ .
\label{eq:FIandDilIntegral}
\ee
The Hubble parameter $H(a^\prime)$ is given by the Friedmann equation
\be
H(a^\prime) = \left( \frac{8 \pi}{3} \right)^{1/2} \frac{\sqrt{\rho_R(a^\prime) + \rho_{NR}(a^\prime)}}{M_{Pl}} \ ,
\label{eq:H}
\ee
where $\rho_{NR}$ is the non-radiation component of the energy density and the temperature is related to the radiation energy density $\rho_R$ by \Eq{eq:rhoR}.

For late times the comoving density $\mathcal{X}$ approaches a constant value $\mathcal{X}_\infty$, and this value can be used to compute the DM density today. At $T_R$ the universe enters the usual RD era and we can trust entropy conservation again. Therefore, we identify a late temperature less than $T_R$ and evaluate $\rho_X$ and the entropy $s a^3$. This ratio in the subsequent evolution of the universe must be constant. We define
\be
\xi_X = \frac{\rho_X(T < T_R)}{s(T < T_R)} = m_X \frac{n_X(T<T_R)}{s(T < T_R)} = 
m_X \frac{\mathcal{X}(T < T_R)}{S(T < T_R)} = m_X \frac{\mathcal{X}_\infty}{S(T < T_R)}   \ .
\label{ed:xiXFIandD}
\ee
The energy density today reads
\be
\rho_X(t_0) = \xi_X  \, s_0 \ , \qquad \qquad \qquad s_0 = 2891 \, {\rm cm}^{-3} \ .
\ee
The contribution to the dark matter energy density reads
\be
\Omega_X h^2 = \frac{\rho_X}{\rho_{\rm cr} / h^2} \ , \qquad \qquad \qquad 
\rho_{\rm cr} / h^2 = 1.054 \times 10^{-5} \,  \GeV \, {\rm cm}^{-3} \ .
\ee

This solution is completely general and valid for Freeze-In during an arbitrary background of radiation and non-radiation components. 

\section{Irrelevance of LOSP Freeze-Out in MD Era}
\label{sec:FO}

So far we have not discussed the LOSP Freeze-Out (FO) contribution to the DM density. If we assume the LOSP particle $B$ to be a generic particle with a weak scale mass and couplings, and in the absence of dilution, this contribution is in the right ballpark to account for the observed DM abundance. The goal of this Section is to show that, in the displaced signal regions for FI identified above, FO is always suppressed and can be safely neglected.

The LOSP FO calculation can be performed following the strategy of Refs.~\cite{Chung:1998rq,Giudice:2000ex}. The Boltzmann equation describing the evolution of the number of $B$ particles reads
\be
\frac{d n_B}{d t} + 3 H n_B = - \langle \sigma v_{{\rm rel}} \rangle_{B B \rightarrow {\rm all}} \left[n_B^2 - n_B^{{\rm eq}\,2}\right] \ .
\label{eq:BoltzFO}
\ee
We parameterize the annihilation cross section as follows
\be
\langle \sigma v_{\rm rel} \rangle_{B B \rightarrow {\rm all}}  \simeq \frac{\alpha_B^2}{32 \pi m_B^2} \ .
\ee
As shown in \App{sec:FIsol}, when there is dilution the variable $Y_B$ is not convenient, since entropy is not conserved. Analogously to the previous case we define $\mathcal{B} = n_B a^3$ and the Boltzmann equation~(\ref{eq:BoltzFO}) in terms of the new variable $\mathcal{B}$ and with time evolution accounted for by the scale factor results in
\be
\frac{d \mathcal{B}}{d \ln a} = - \frac{\langle \sigma v_{{\rm rel}}\rangle}{H(a) a^3}  \left[\mathcal{B}^2(a) - n_B^{\rm eq}(T(a))^2 \, a^6\right]   \ ,
\label{eq:FOandDilution}
\ee
with the Hubble parameter as in \Eq{eq:H}. For large values of the scale factor, corresponding to temperature much greater than $T_R$, the variable $\mathcal{B}$ approaches an asymptotic value $\mathcal{B}_\infty$. The present energy density of $X$ particles is computed analogously to the Freeze-In case
\be
\xi_X = m_X \frac{\mathcal{B}_\infty}{S(T < T_R)}   \ .
\label{ed:xiXFOandD}
\ee
Notice the difference with the Freeze-In case in \Eq{ed:xiXFIandD} where $\mathcal{B}_\infty$ is replaced by $\mathcal{X}_\infty$.

In what follows we consider different cosmologies and different values of $m_B$ and $m_X$, and for each case we set $\alpha_B$ in such a way that in the absence of dilution \Eq{ed:xiXFOandD} reproduces the observed DM density. We find the following values
\be
\alpha_B = \left\{ \begin{array}{lccl} 
0.07  & & & (m_B, m_X) = (300 \GeV, 100 \GeV)    \\
0.7  & & & (m_B, m_X) = (3 \TeV, 1 \TeV) 
\end{array} \right.  \ .
\label{ed:AlphaBValues}
\ee
Then we explore the impact of dilution on these reference points.  

\begin{figure}
\begin{center}
\includegraphics[scale=0.4]{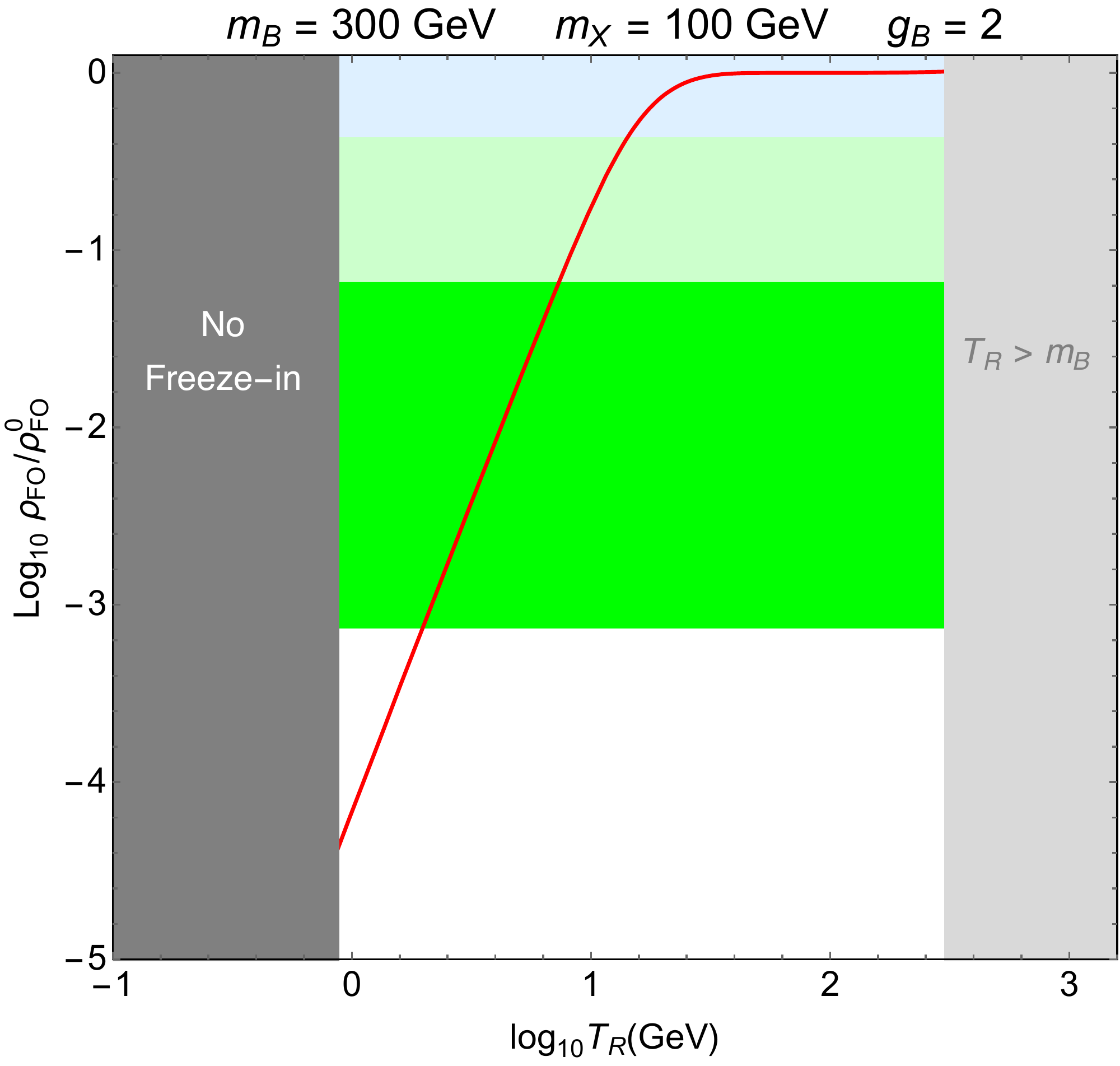} $\quad$
\includegraphics[scale=0.4]{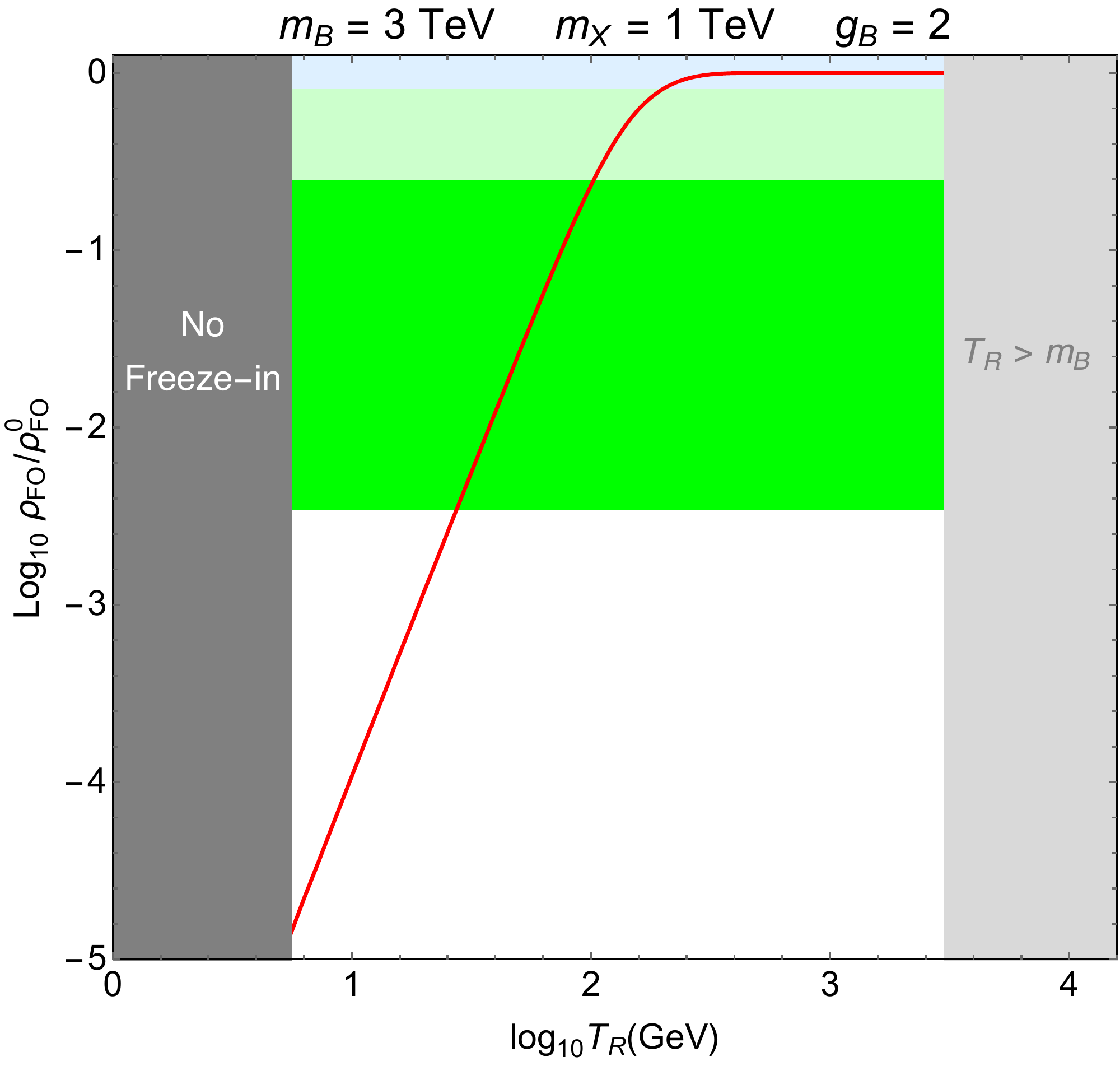} 
\end{center}
\caption{Suppression of LOSP FO contribution to the $X$ relic density as a function of $T_R$ for inflationary reheating. We show two different choices of $(m_B, m_X$) masses. Displaced collider signals occur in the shaded horizontal bands as described in Table~\ref{table:DisplacedColliderSignals}.}
\label{fig:FreezeOutDilutionINF}
\end{figure}

\begin{figure}
\begin{center}
\includegraphics[scale=0.4]{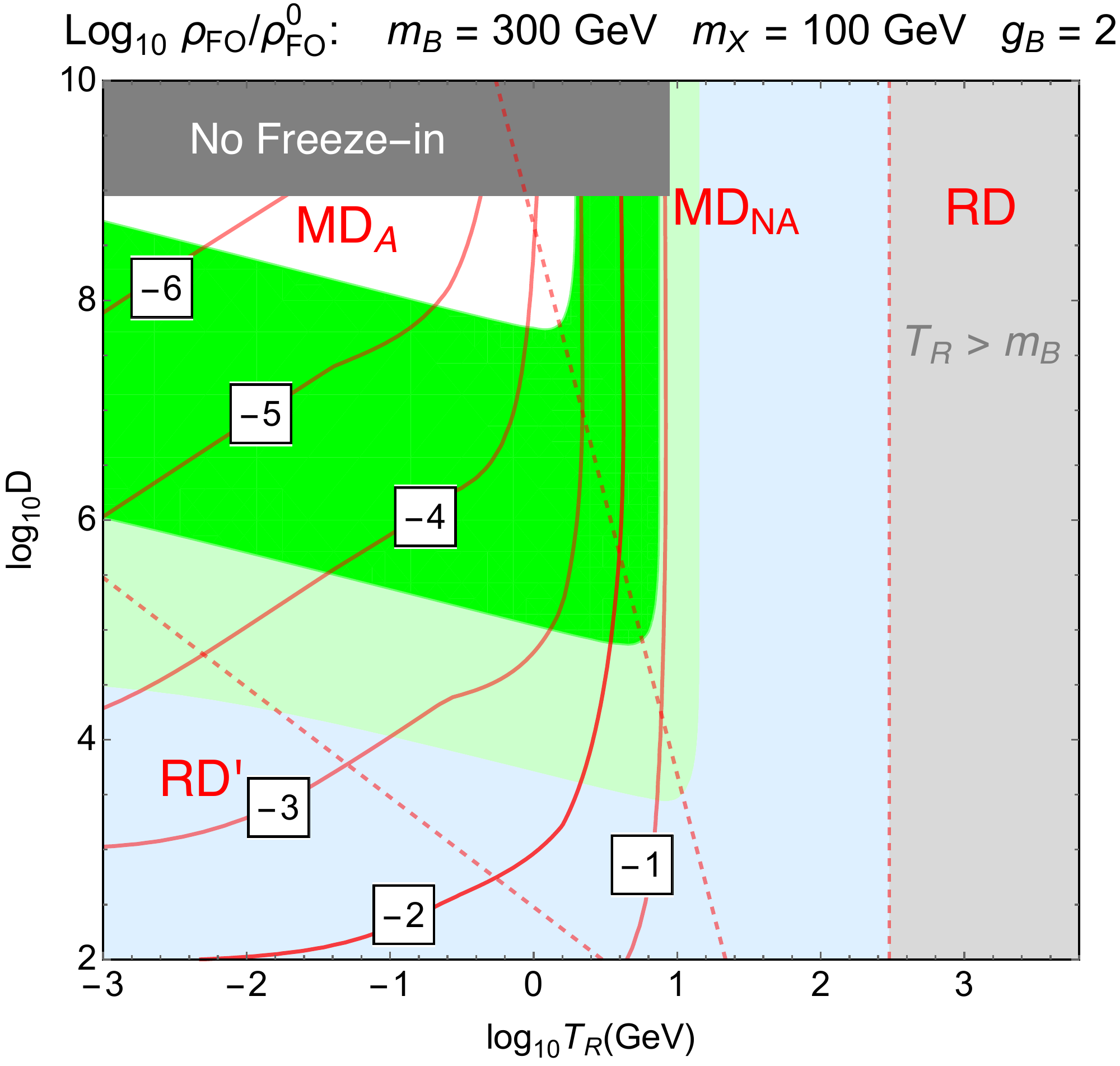} $\quad$
\includegraphics[scale=0.4]{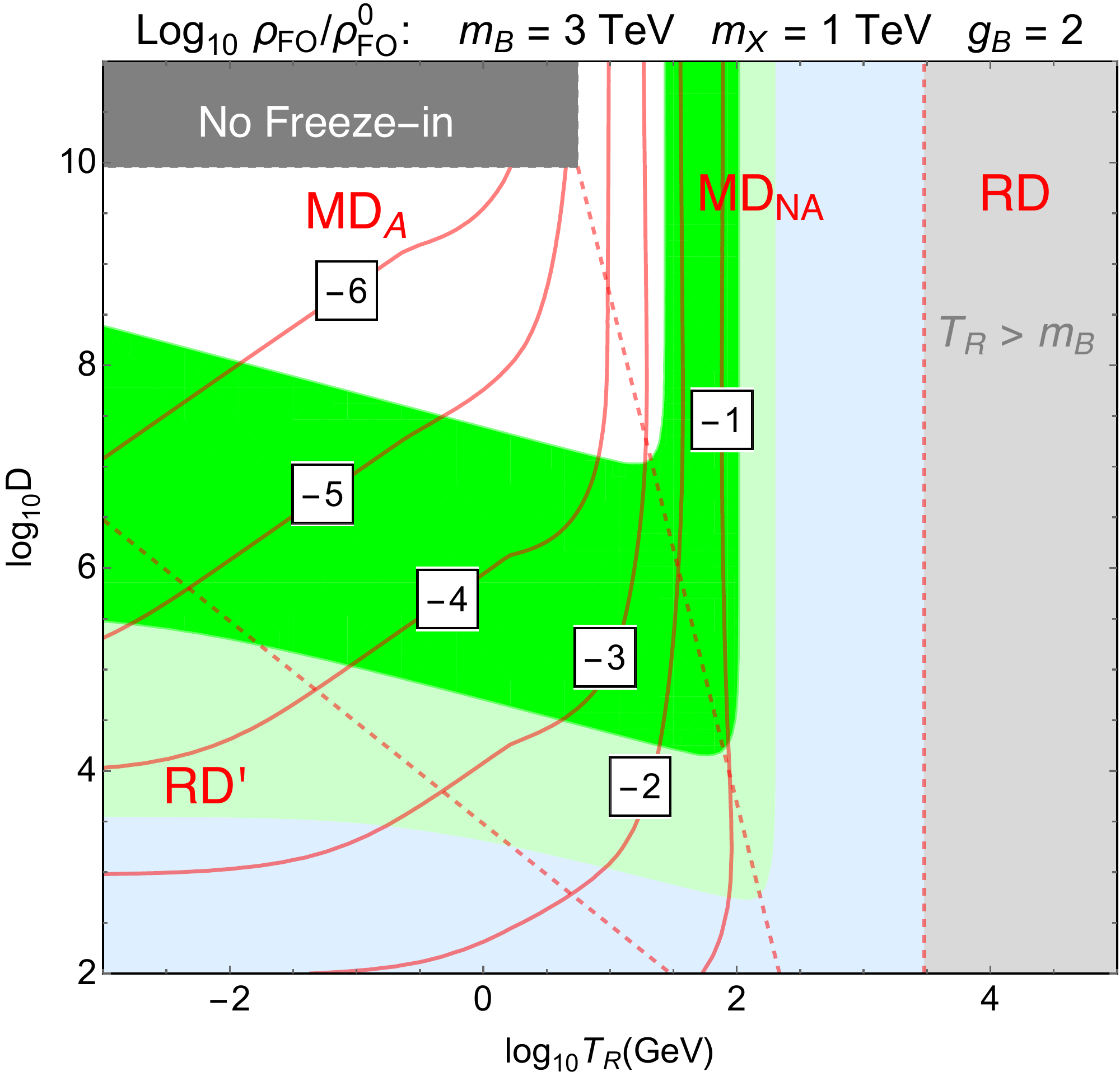}
\end{center}
\caption{Suppression of LOSP FO contribution to the $X$ relic density in the $(T_R, D)$ plane for two different choices of $(m_B, m_X$);  for lighter $X$ the suppression will be greater. Displaced collider signals occur in the green and blue shaded bands, as described in Table~\ref{table:DisplacedColliderSignals}.}
\label{fig:FreezeOutDilution}
\end{figure}

We show our results for the case of inflationary reheating in Fig.~\ref{fig:FreezeOutDilutionINF}.  The red curve gives the dilution factor as a function of $T_R$ over the range of $T_R$ where FI can occur, with the two panels corresponding to the two cases of \Eq{ed:AlphaBValues}. The shaded regions denote the associated collider signal for the chosen masses and for the value of $T_R$ on the red curve, as found in Fig.~\ref{fig:MasterPlotINF}.


Next we consider the more general case of both $MD_{\rm A}$ and $MD_{\rm NA}$, with results in Fig.~\ref{fig:FreezeOutDilution} for the same choices of $B$ and $X$ masses. The contours of both panels show that the FO contribution is always reduced as either $T_R$ is reduced or $D$ is increased.  The shaded regions of both panels refer to the FI contribution, and are taken from the appropriate panels of Figs. \ref{fig:MasterPlots} and \ref{fig:MasterPlots2}.  In the green regions, where FI gives displaced vertices, the FO contribution is suppressed by 1 to 6 orders of magnitude compared to conventional FO during the RD era, and hence is typically expected to be negligible.

In the right panel of Fig.~\ref{fig:FreezeOutDilution}, the three dashed red lines separate the regions where FO occurs, from right to left, in the (RD, MD$_{NA}$, MD$_A$, RD$'$) eras.  These three boundaries, from right to left, correspond to FO temperatures $T_F$ of $T_R, T_{NA}$ and $T_M$.  The FO abundance in region $i$ is
\be
Y_{\rm{FO}_i} \, \sim \,  \frac{1}{M_{Pl} \, \langle \sigma v \rangle \, T_F}  
 \left( 1, \;\;\; \frac{T_R^3}{T_F^3}, \;\;\; \frac{T_R}{T_M^{1/2} T_F^{1/2}}, \;\;\;\frac{T_R}{T_M} \right)
\label{eq:YFO}
\ee
which can be compared to the FI result of Eq. (\ref{eq:Y03}) and gives the asymptotic slopes of the contours in both panels of Figs.~\ref{fig:FreezeOutDilutionINF} and ~\ref{fig:FreezeOutDilution}.

\end{document}